\documentclass[preprint,showpacs,preprintnumbers,amsmath,amssymb]{revtex4}
\usepackage{graphicx}
\usepackage{dcolumn}
\usepackage{bm}

\begin{document}

\title{Electromagnetic pion and kaon form factors\\in light-cone
resummed perturbative QCD}

\author{Udit Raha}
\email{Udit.Raha@unibas.ch}
\author{Andreas Aste}
 \email{Andreas.Aste@unibas.ch}
\affiliation{Department of Physics, University of Basel, Switzerland.}

\date{\today}


\begin{abstract}
\noindent We analyze the electromagnetic pion and kaon form factor by including
radiative and higher-twist effects within the framework of resummed pQCD in the
space-like region. We focus on the transition from the perturbative to
non-perturbative behavior in the phenomenological intermediate energy regime. 
Using a modified ``$k_T$'' factorization scheme with transverse degrees of 
freedom, we evaluate the non-perturbative soft contributions as distinct from 
the hard contributions, ensuring no double counting via the Ward 
identity at $Q^2\!=\!0$. The soft contributions are obtained via local 
quark-hadron duality while the hard contributions rest on the well known 
collinear factorization theorem using model wave functions with modified 
Brodsky-Huang-Lepage type ansatz and distribution amplitudes derived 
from light-cone QCD sum rules. Our analysis shows that the perturbative hard 
part prevails for large $Q^2$ beyond 50-100 GeV$^{2}$, while for low and 
moderate momentum transfers below 10-16 GeV$^{2}$ the soft contributions 
dominate over the hard part. Thus, we demonstrate the importance of including 
the soft contributions for explaining the experimental form-factor data.
\end{abstract}


\maketitle


\section{Introduction}
\label{sec:intro}
During the past decade, QCD-oriented studies have been shifting steadily
toward exclusive channels. For a long time the electromagnetic
structure of pions has been subjected to numerous experimental and theoretical 
investigations through the study of electroproduction reactions. Extensive 
experimental studies of pion electroproduction reactions like $ep\!\rightarrow 
\!e\pi^+ n$ or $en\!\rightarrow\!e\pi^- p$ have been carried out in the past 
at CERN, Cornell, DESY and more recently at JLab Facility \cite{pion_exp1,pion_exp2,pion_exp3,pion_exp4,pion_exp5,pion_exp6,pion_exp7,pion_exp8}. 
Since the mid-90s, kaon electroproduction reactions like $A(\gamma,K)YB$ and 
$A(e,e'K)YB$ ($A$ is the target, $Y$ the produced hyperon and $B$ the recoil) 
have also attracted renewed interest in nuclear physics at both experimental 
\cite{kaon_exp1,kaon_exp2} and theoretical \cite{kaon_theo1,kaon_theo2} level. 
The main ingredients for the description of electromagnetically induced kaon 
production are embedded in the so-called Chew, Goldberger, Low and Nambu (CGLN)
scattering amplitudes. In case of the longitudinal component of the electron 
induced unpolarized differential cross-section, the t-channel diagram dominates
and (in certain kinematic conditions) can be factorized
\cite{kaon_exp1,Williams} as $\sigma_L=k\cdot{\mathcal F}(Q^2)\,{\mathcal
  G}(W){\mathcal H}(t)$, where $k$ is a kinematic factor, ${\mathcal F}$,
${\mathcal G}$, and ${\mathcal H}$ are functions of the 4-momentum transfer
squared $Q^2$ of the virtual photon, the invariant mass $W$ and the Mandelstam 
variable $t$, respectively. The function ${\mathcal F}(Q^2)$ implicitly 
contains the information about the electromagnetic form factor of the kaon. 
Note that ${\mathcal F}(Q^2)$ is not the actual form factor, but rather a 
complicated function from which the form factor can be extracted using e.g., 
Chew-Low extrapolation and deconvolution algorithms. A precise knowledge of the
form factor is of fundamental importance for a realistic and accurate 
description of {\it exclusive reaction} mechanisms and plays a key role 
in understanding the interplay between perturbative and non-perturbative 
physics at intermediate energies. Moreover, the study of form factors 
provides direct insight into the electromagnetic structures and charge 
distributions of hadrons as they couple with photons.

To date, the electromagnetic kaon form factor is very poorly known and only 
measured at very low $Q^2$ (below $0.2$ $\mbox{GeV}^2$) \cite{pion_exp5,Dally}.
 The status for the (quasi-free) Lambda ($\Lambda$) and Sigma ($\Sigma$) 
hyperons is even worse, i.e., there are simply no available experimental data. 
Basic quantities like the strong coupling constants $g_{K \Lambda N}$ and 
$g_{K \Sigma N}$ derived from purely hadronic processes or theoretical 
considerations are not well established and must be considered adjustable. 
Recently, however, there appeared quite large and precise data sets on 
photo-production of kaons from the SAPHIR (ELSA) \cite{Glander}, CLAS (CEBAF) 
\cite{McNabb} and LEPS (SPring8) \cite{Zegers} collaborations. There is also 
new data on electroproduction of positive kaons from experiment E98-108 at 
CEBAF which are being analyzed at the moment.
 
Keeping in mind the increasing accuracy of experimental data, an 
accurate theoretical description of the electromagnetic form factors of
pseudo-scalar charged mesons at intermediate energies is of primal importance. 
To our knowledge, especially for the kaon there are very few theoretical works 
in this direction \cite{kff1,kff2,kff3,Bijnens}. In this paper, in addition to 
the pion form factor, we analyze the kaon form factor for a broad range of 
space-like momentum transfer. Our framework is based on resummed perturbative 
light-cone QCD formalism \cite{St,LiSt}, unlike conventional approaches like 
``asymptotic'' and lattice QCD or from sum rules that rely on many unchecked 
hypotheses. The experimental results could then be used to extract the various 
{\it distribution amplitudes} (DAs) \cite{wfs0,wfs1,wfs2,wfs3,wfs4,wfs5,wfs6,wfs7,wfs8,wfs9,DAs1,DAs2,DAs3,DAs4,DAs5,frozen1,frozen21,frozen22}
used in the above formulation. Of course, only a handful of experimental
hadron electroproduction data points are presently available to make definitive
statements on the validity of different theoretical approaches. Since in almost
all cases, the corresponding data points are merely concentrated in 
the very low energy region ($Q^2<1$ GeV$^2$), perturbative QCD (pQCD) has 
limited predictive power due to the rapidly growing magnitude of the strong 
coupling, as $Q^2$ tends to zero. Despite the existing plethora of literature 
on the predictions on electromagnetic meson form factors based on various 
approaches (see e.g., 
Refs.~\cite{Bijnens,St,LiSt,frozen1,frozen21,frozen22,VMD1,VMD2,VMD3,Anant,asy,FJ,pff01,pff02,pff03,pff04,pff11,pff12,pff13,pff14,pff15,pff2,pff3,pff4,pff5,pff6,pff7,pff8,pff9,Lattice1,Lattice2,Lattice3,Lattice4,crit4,Brodsky,Efremov,Radyushkin}
for the pion and Refs.~\cite{kff1,kff2,kff3,Bijnens} for kaon form factors, to 
give a highly incomplete list of references), till date there is considerable 
amount of debate as to their exact behavior in the phenomenological low and 
moderate energies between $Q^2\approx 4$-$50$ GeV$^2$. Nevertheless, we try to 
give our assessment to the existing scenario and try to explain the 
experimental data first for the pion form factor, where statistics are far more
decent as compared to that of the kaon. Then we extend our analysis to the kaon
form factor, where experimental data is still too limited for any meaningful 
comparison. Hopefully, with the planned 12 GeV upgrade proposal of the CEBAF 
experiment (at JLab) in the near future, studies of intermediate energy QCD 
can prove to be fruitful.

The standard ``asymptotic'' QCD is known to make successful 
predictions of many phenomena like {\it dimensional scaling, helicities, color 
transparency}, etc.~for exclusive processes, as $Q^2$ tends to infinity 
\cite{FJ,Brodsky,Efremov,Radyushkin,Mueller}. The approach relies on the 
so-called {\it collinear factorization} theorem \cite{Efremov,Radyushkin} which
provides an outstanding way of isolating the partonic part accessible to pQCD 
from the non-perturbative parts. The basic ingredients are: (a) the hadron DA 
$\phi$, which encodes the non-perturbative information regarding the momentum 
distribution of the constituent ``near'' on-shell valence partons collinear to 
the hadron and also features of the QCD vacuum structure as expressed through 
the quark condensates \cite{condensate1,condensate2,condensate3}, and (b) a 
scattering kernel $T_H$, describing the hard scattering of ``far'' off-shell 
valence partons. The overall amplitude of the exclusive process is then given 
by the convolution, $\phi\otimes T_H \otimes\phi$. However, the application of 
pQCD to exclusive processes at intermediate momentum transfers or 
phenomenologically accessible energies (e.g., at CEBAF) has been the subject of
severe controversies and criticisms \cite{crit4,crit0,crit1,crit2,crit3,crit5,crit6,crit7,crit8,crit9}. It is 
widely anticipated that non-perturbative effects arising from soft gluon 
exchanges or from endpoint contributions to phenomenologically acceptable wave 
functions (or DAs) dominate and may severely preclude the predictability of 
pQCD. Hence, in this paper we use a modified ``resummed'' pQCD formalism
(as proposed in Refs.~\cite{St,LiSt}) which is believed to largely enhance the
predictability of pQCD in a self-consistent way at intermediate energies. The 
central issue here is the inclusion of transverse momentum $k_T$ dependence 
that necessitates the inclusion of a {\it Sudakov suppression} factor. This is 
to organize the large double logarithms of the type $\alpha_s\ln^2 k_T$, 
arising at all orders due to the overlap of soft and collinear contributions 
of radiative gluon loop corrections. Such a resummation effectively suppresses
the non-perturbative contributions at large energies. However, this may not 
still be effective enough when talking of $Q^2$ down to a few GeV$^2$. These 
facts are in agreement with some of the recent findings, reported in 
Refs.~\cite{pff7,pff8,pff9} for the pion form factor and also in the context 
of $B$ systems \cite{Descotes}.
         
In this paper, we emphasize the importance of including two distinct
contributions to exclusive quantities for obtaining good agreement with 
experimental data at low and moderate energies: firstly, the non-factorizable 
soft contributions which are not calculable within the perturbative framework
and secondly, the power suppressed corrections from non-leading twist 
structures (twist-3) determining the preasymptotic behavior. In other words, 
the electromagnetic form factor $F_M(Q^2)$ for a charged meson $M$ should be 
written as
\begin{equation}
F_M(Q^2)=F^{\rm soft}_M(Q^2)+F^{\rm hard}_M(Q^2)\,,
\end{equation}
where $F^{\rm hard}_M(Q^2)$ is the factorizable part computable in pQCD 
and $F^{\rm soft}_M(Q^2)$ is the non-factorizable soft part. The soft 
contributions to the form factors can be calculated using phenomenological 
quark models either incorporating transverse structure (momentum) dependence 
of the hadron wave functions (see e.g.,
Refs.~\cite{pff5,crit1,crit3,crit5,BHL}) or from QCD sum rules via {\it Local 
Duality} (see e.g., Refs.~\cite{pff01,pff02,pff03,pff04,pff6,SVZ}.) In the 
present paper, we follow the latter approach. We also focus on the presence of 
non-perturbative enhancements arising from kinematic endpoint regions of the 
scattering kernel which tend to invalidate collinear factorization. For our 
calculations, we use model twist-2 and twist-3 light-cone wave functions 
incorporating transverse degrees of freedom, where the collinear DAs are 
derived from QCD sum rules \cite{DAs1,DAs2,DAs3,DAs4,DAs5}. 
Naively, the twist-3 contributions to the form factor are expected to be small 
compared to leading (twist-2) contributions as they have a relative $1/Q^2$ 
suppression. On the contrary, the existing literature, either using model or 
asymptotic DAs \cite{wfs1,wfs3,wfs4,pff2,pff7,pff9}, shows large twist-3 
corrections to the pion form factor which even overshoot the twist-2 
contributions in a wide range of low and intermediate energies. This is also 
confirmed in our analysis and is in fact more enhanced for the kaon form 
factor. To this end, our analysis shows good agreement with the existing 
pion data and in addition we prove the consistency of our results by adopting a
scheme of analytization of the running strong coupling 
\cite{pff5,pff6,Shirkov,analytic} that removes the explicit Landau singularity 
at $Q^2\!=\!\Lambda^2_{\rm QCD}$ by a minimum power correction in the UV 
regime.

The paper is organized as follows: In Section~\ref{sec:DAs}, we 
briefly discuss the idea of factorization and review the basic definitions of 
the twist-2 and twist-3 pseudo-scalar meson DAs and their renormalization 
evolutions. Section~\ref{sec:form} deals with the theoretical framework 
involved in calculating the space-like electromagnetic form factor. Here, we 
recall the predictions of classic asymptotic QCD for large 
$Q^2\!\rightarrow\!\infty$ and how one needs to modify pQCD with 
collinear as well as ``$k_T$'' factorization schemes including Sudakov effects 
at intermediate energies. In Section~\ref{sec:numerics}, we provide the details
of our numerical results for the pion form factor and compare it with the 
available experimental data. We also give a preliminary prediction for the 
kaon form factor, despite the lack of available experimental data for
comparison in the desired phenomenological regime. Finally, 
Section~\ref{sec:concl} contains our summary and conclusions. The appendices 
contain a compendium of relevant formulae used in our analysis.
 
\section{Factorization and Distribution Amplitudes}
\label{sec:DAs}
The parton model of describing exclusive processes in QCD inherently rest on
the so-called {\it frozen approximation} (\cite{Brodsky,Efremov,Radyushkin}). 
At high energies, exclusive scattering amplitudes are dominated by hadronic 
Fock states with essentially valence quark configurations ($\bar{q}q$ in 
mesons). While the relative velocities of the participating hadrons are located
closely to the null-plane, the internal hadron ``quantum-bindings'' processes 
are highly time-dilated with respect to the exclusive reaction time scales in 
the rest frames of the remaining hadrons. This effectively freezes the hadronic
internal degrees of freedom as seen by the other hadrons. This incoherence 
between the long-distance intra-hadronic binding processes and the 
short-distance inter-hadronic scattering reaction is the very motivation for 
the idea of factorization. Thus, the hadrons may be considered to be consisting
of definite valence quark states denoted by a DA of leading twist $\phi$. The 
collinear factorization formula is then used to express exclusive quantities 
like the form factors as a convolution using the DAs:
\begin{equation}
\label{eq:factor}
F_M(Q^2)=\int^{1}_{0}dx dy \,\phi_{\rm
in}(x,\mu^2_{\rm F})\,T_H(x,y,Q^2,\mu^2_{\rm F},\mu^2_{\rm R})\,\phi_{\rm out}
(y,\mu^2_{\rm F})+\cdots\,,
\end{equation}
where $Q^2=-2P_{\rm in}\cdot P_{\rm out}$. Here, $P_{\rm in}$ and $P_{\rm out}$
are, respectively, the ingoing and outgoing hadron momenta, $x$ and $y$ are the
longitudinal momentum fractions of the nearly on-shell valence quarks, 
$\mu_{\rm R}$ is the renormalization scale and  $\mu_{\rm F}$ is the 
factorization scale which is defined as the scale below which the QCD dynamics 
are non-perturbative and remain implicitly encoded within the DAs, while the 
dynamics above are perturbative and must be retained in the hard kernel $T_H$. 
The ellipses in the above equation represent contributions from higher 
order Fock states and sub-leading twists which are all suppressed by inverse 
powers of $Q^2$. In addition, they also include the non-factorizable soft 
contributions. Formally, the definition of the leading twist-2 DA for 
pseudo-scalar mesons (e.g., $\pi^-$) can be given in a process- and 
frame-independent manner \cite{DAs1,DAs2,Brodsky,Efremov,Radyushkin} in terms 
of matrix elements of non-local light-ray operator along a certain light-like 
direction $z_\mu \,(z^2=0)$: 
\begin{equation}
\left<0\left|\bar{u}(z)[z,-z]\gamma_\mu \gamma_5 d(-z)\right|\pi^-(P)\right> =
iP_\mu\,\int^1_0 dx\, e^{i\xi(zp)}\phi_{2;\pi}(x,\mu^2_{\rm F})\,;\, \xi=2x-1
\,,
\end{equation}
with the path-ordering (${\mathcal P}$) Wilson line in terms of the gluon
``connection'' along the straight line joining $z$ and $-z$ along the 
null-plane given by
\begin{equation}
\label{eq:Wilson}
[z,-z]={\mathcal P}\left[ig_s\int^z_{-z}dy^{\,\mu} A_\mu (y)\right]\,,
\end{equation}
where $P^2_\mu=m^2_\pi$ and $p_\mu$ is a light-like vector,
\begin{equation}
p_\mu=P_\mu-\frac{1}{2}z_\mu\frac{m^2_\pi}{Pz}\,.
\end{equation}
The local limit $z\rightarrow 0$ gives the normalization condition at an
arbitrary scale $\mu$,
\begin{equation}
\label{eq:norm-twist2pi}
\int^1_0\,\phi_{2;\pi}(x,\mu^2)\, dx = \frac{f_\pi}{2\sqrt{2N_c}}
\end{equation}        
with the pion decay constant, $f_\pi\approx 131$ MeV defined by
\begin{equation}
\left<0\left|\bar{u}(0)\gamma_\mu \gamma_5 d(0)\right|\pi^-(P)\right> =if_\pi 
P_\mu\,.
\end{equation}
The leading twist-2 DA $\phi_{2;\pi}(x,\mu^2)$ can be expressed as a conformal 
series expansion over Gegenbauer polynomials 
$C^{\,3/2}_{2n}$ :
\begin{equation} 
\label{eq:pion-DA2}
\phi_{2;\pi}(x,\mu^2)=\frac{3f_\pi}{\sqrt{2N_c}}\,x(1-x)\left(1+\sum^{\infty}_{n=1}
  a^\pi_{2n}(\mu^2)\,C^{3/2}_{2n}(\xi)\right)\,,
\end{equation}
where
\begin{equation}
\label{eq:asy}
\phi^{({\rm as})}_{2;\pi}(x)=\phi_{2;\pi}(x,\mu^2\rightarrow \infty)=\frac{3f_\pi}{\sqrt{2N_c}}\,x(1-x)
\end{equation}
is generally referred to as the {\it asymptotic} DA. The Gegenbauer moments 
$a^\pi_{2n}$ represent the non-perturbative inputs encoding the long-distance 
dynamics and may be obtained e.g., via lattice QCD calculations or QCD sum 
rules. The renormalization group (RG) equation for $\phi_{2;\pi}(x,\mu^2)$ is 
known as the Efremov-Radyushkin-Brodsky-Lepage (ER-BL) equation 
\cite{Brodsky,Efremov,Radyushkin},
\begin{equation}
\mu^2\frac{d}{d\mu^2}\phi_{2;\pi}(x,\mu^2)=\int^1_0 dy\,V(x,y;\alpha_s(\mu^2))\,
\phi_{2;\pi}(y,\mu^2)
\end{equation}
with the integral kernel $V(x,y;\alpha_s)$ to leading order in $\alpha_s$ 
given by
\begin{equation}
V_0(x,y;\alpha_s)={\mathcal C}_F\frac{\alpha_s}{2\pi}\left[\frac{1-x}{1-y}\left(1+
\frac{1}{x-y}\right)\theta(x-y)+\frac{x}{y}\left(1+\frac{1}{y-x}\right)\theta
(y-x)\right]_+\,,
\end{equation}
where the ``+'' distribution is defined as
\begin{equation}
\left[V(x,y;\alpha_s)\right]_+=V(x,y;\alpha_s) - \delta(x-y)\int^1_0 dt\,V(t,y;\alpha_s)\,.
\end{equation}
Solving the above set of equations yields the multiplicative renormalization
formula for moments $a^\pi_n$ to leading-logarithmic accuracy,
\begin{equation}
a_n(\mu^2)=L^{\gamma^{(0)}_n/\beta_0}a_n(\mu^2_0)\,,
\end{equation}
where $L=\alpha_s(\mu^2)/\alpha_s(\mu^2_0)$ and $\beta_0=(11N_c-2N_f)/12$,
while the lowest order anomalous dimensions are given 
by
\begin{equation}
\gamma^{(0)}_n={\mathcal C}_F\left(\psi(n+2)+\psi(1)-\frac{3}{4}-\frac{1}
{2(n+1)(n+2)}\right)
\end{equation}
with the logarithmic derivative of the Gamma function 
$\psi(z)=\Gamma'(z)/\Gamma(z)$.
Note that for the pion all odd moments $a^\pi_{n=1,3,5\cdots}$ vanish due to 
isospin symmetry. In contrast, the kaon DA have non-zero values for the odd 
moments signifying flavor-SU(3) violation effects. Hence, the twist-2 DA for 
the kaon (e.g., $K^-$) is given by the expansion
\begin{eqnarray} 
\label{eq:kaon-DA2}
\phi_{2;K}(x,\mu^2)=\frac{3f_K}{\sqrt{2N_c}}\,x(1-x)\!\!\!\!&&\!\!\!\!\!\!\left(1+\sum^
{\infty}_{n=1}a^K_n(\mu^2)\,C^{3/2}_n(\xi)\right);\\
\int^1_0\,\phi_{2;K}(x,\mu^2)\, dx\!\! &=& \!\!\frac{f_K}{2\sqrt{2N_c}}\,,
\end{eqnarray}
where the kaon decay constant $f_K\approx 1.22f_\pi$ \cite{wfs0} is defined 
by
\begin{equation}
\left<0\left|\bar{u}(0)\gamma_\mu \gamma_5 s(0)\right|K^-(P)\right> =if_K 
P_\mu\,.
\end{equation}
The Gegenbauer moments being multiplicatively renormalizable with growing 
anomalous dimensions, for sufficiently large renormalization scale a 
finite number of moments are relevant, albeit the fact that the higher order 
moments have large uncertainties in their present determination. Hence, in 
all practical calculations, the series expansion of the DAs are truncated only 
to the first few moments. In this paper, we have adopted a model for the 
twist-2 DAs in truncating up to the second moment, as was done in 
Refs.~\cite{DAs2,DAs4}.

For the charged pseudo-scalar mesons at the twist-3 level, there are 
two 2-particle DAs and one 3-particle DA. Here, we only give the formal 
definitions of the 2-particle DAs that we need in our analysis. For the 
charged pion (e.g., $\pi^-$), they are defined by \cite{DAs1} :
\begin{eqnarray}
\left<0\left|\bar{u}(z)\,i\gamma_5\, d(-z)\right|\pi^-(P)\right>\!\! &=&
\!\!\mu_\pi\int^1_0 dx\,e^{i\xi(zp)}\phi^{\,p}_{3;\pi}(x,\mu^2)\,,\nonumber\\
\left<0\left|\bar{u}(z)\,\sigma_{\alpha\beta}\gamma_5\, d(-z)\right|\pi^-(P)
\right>\!\!&=&\!\!-\frac{i}{3}\,\mu_\pi(P_\alpha z_\beta -P_\beta z_\alpha)\,\int^1_0 dx\,e^{i\xi(zp)}\phi^\sigma_{3;\pi}(x,\mu^2)
\end{eqnarray}
with $\mu_\pi=m^2_\pi/(m_u+m_d)$ and similarly for the charged kaon 
(e.g., $K^-$) \cite{DAs4} :
\begin{eqnarray}
\left<0\left|\bar{u}(z)\,i\gamma_5\, s(-z)\right|K^-(P)\right>\!\! &=&\!\!
\mu_K\int^1_0 dx\,e^{i\xi(zp)}\phi^{\,p}_{3;K}(x,\mu^2)\,,\nonumber\\
\left<0\left|\bar{u}(z)\,\sigma_{\alpha\beta}\gamma_5\,s(-z)\right|K^-(P)\right>\!\!&=&\!\!-\frac{i}{3}\,\mu_K(P_\alpha z_\beta -P_\beta z_\alpha)\,\int^1_0 dx\,e^{i\xi(zp)}\phi^\sigma_{3;K}(x,\mu^2)
\end{eqnarray}
with $\mu_K=m^2_K/(m_u+m_s)$. Note that the gauge-link factors (Wilson 
line (\ref{eq:Wilson})) in the matrix elements are to be implicitly 
understood. The twist-3 DAs have the following asymptotic forms:
\begin{eqnarray}
\label{eq:twist-3asy}
\phi^{{p}\,({\rm as})}_{3;M}(x)\!\!&=&\!\!\frac{f_{M}}{4\sqrt{2N_c}}\,,
\nonumber\\ 
\phi^{{\sigma}\,({\rm as})}_{3;M}(x)\!\!&=&\!\!\frac{3f_{M}}{2\sqrt{2N_c}}\,x
(1-x)\,\,;\,\,M=\pi^\pm,K^\pm
\end{eqnarray}
with the normalization condition,
\begin{equation}
\label{eq:norm-twist3}
\int^1_0\,\phi^{\,p,\sigma}_{3;M}(x,\mu^2)=\frac{f_{M}}{4\sqrt{2N_c}}
\,.
\end{equation}
For our analysis, we use the 2-particle twist-3 DAs from Refs.~\cite{DAs2,DAs4}
defined at the scale $\mu=1$ GeV. As a matter of book-keeping, we explicitly 
provide the relevant formulae for the charged pion and kaon DAs in 
Appendix~\ref{app:DA}.

\section{Space-like electromagnetic form factor}
\label{sec:form}
The electromagnetic form-factor is considered as an important observable for
studying the onset of the perturbative regime in exclusive processes. For large
$Q^2$, the asymptotic scaling behavior $ F_M(Q^2)\sim 1/Q^2$ follows from the
well-known dimensional ``quark counting'' while for small $Q^2$, the behavior
is well described by the Vector Meson Dominance (VMD) model \cite{VMD1,VMD2,VMD3} and given by
\begin{equation} 
 F_M(Q^2)\approx \frac{1}{1+Q^2/\mu^2_{\rm VDM}}\,\,;\,\,Q^2\ll\mu^2_{\rm VDM} \, ,
\end{equation}
where $\mu_{\rm VDM}\approx 750$ MeV is a reasonable cut-off mass scale,
showing no obvious trace of pQCD scaling behavior where there exists no high 
energy cut-off. Hence, a thorough understanding of this transition behavior
(from nonperturbative to perturbative) is of crucial importance in QCD for 
understanding the very nature of strong interaction and in providing a vivid 
picture of the underlying quark-gluon substructure of the mesons.

For a charged meson $M$ (e.g., $\pi^\pm, K^\pm$), the form factor is 
specified by the following matrix element:
\begin{equation}
(P '+P)_\mu\, F_M(Q^2)=\left<M(P ')\left|J_\mu(0)\right|M(P)\right>\,\,;
\,\,J_\mu=\sum^{\,}_{f}e_f\bar{q}_f\gamma_\mu q_f\,,
\end{equation}
where $J_\mu$ is the electromagnetic current with quark $q_f$ of flavor $f$
and charge $e_f$. In this paper, we shall only consider space-like momentum
transfers i.e., $q^2=(P'-P)^2=-Q^2$. Neglecting the meson masses, we consider 
the ``brick wall'' frame where the incoming particle with 4-momentum $P$ in the
$z$ direction recoils with 4-momentum $P'$ in the $-z$ direction after
interacting with the hard photon ``wall''. In the light-cone formalism, 
$P=(Q/\sqrt{2},0,{\bf 0}_{T})$ and $P'=(0,Q/\sqrt{2},{\bf 0}_{T})$.

\subsection{Hard contribution in pQCD}
The hard contributions to the form factor are calculated using the collinear
factorization formula Eq.~(\ref{eq:factor}), where the hard scattering kernel 
$T_H$ at the scale $\mu=\mu_{\rm _F}=\mu_{\rm R}$ is given to the
leading order in $\alpha_s$ by
\begin{equation}
T_H(x,y,Q^2,\mu^2)=16\pi {\mathcal C}_F\,\alpha_s(\mu^2)\left[\frac{2}{3}
\frac{1}{xyQ^2}+\frac{1}{3}\frac{1}{(1-x)(1-y)Q^2}\right]\,,\end{equation}
where in QCD the value of the {\it Casimir} \,operator in the fundamental 
representation of SU(3) is ${\mathcal C}_F=(N^2_c-1)/2N_c=4/3$. The 
factorization formula then yields the classic pQCD expression for the meson 
form factor at $\mu^2=Q^2$ :
\begin{equation}
F^{\rm hard}_M(Q^2)=\frac{16\pi\,{\mathcal C}_F\,\alpha_s(Q^2)}{Q^2}\left|\int^1_0 dx \frac{\phi_{2;M}(x,Q^2)}{x}\right|^2\,.
\end{equation}
Note that using the asymptotic twist-2 DA $\phi^{({\rm as})}_{2;M}(x)$, one
obtains the familiar $1/Q^2$ scaling behavior for $Q^2\rightarrow\infty$ 
\begin{equation}
\label{eq:hard-asy}
F^{\rm hard}_M(Q^2)=\frac{8\pi\alpha_s(Q^2)f^2_M}{Q^2}\,.
\end{equation}
The principal motivation of the modified ``resummed'' pQCD is the elimination 
of large logarithms in the hard kernel that arise from radiative gluon loop 
corrections. One way of doing this is by the introduction of intrinsic 
transverse momenta dependence of the constituent partons, giving rise to a 
Sudakov suppression due to certain partial resummation of transverse terms, as 
mentioned earlier in the introduction. Including the transverse momenta of the 
two valence quarks within the meson, the tree-level hard kernel $T_H$ in the 
momentum-space is written as
\begin{equation}
T_H(x,y,Q^2,{\bf k}_{1T},{\bf k}_{2T},\mu^2)=\frac{16\pi\,{\mathcal
C}_F\,\alpha_s(\mu^2)\,xQ^2}{(xQ^2+{\bf k}_{1T}^2)(xy\,Q^2+(
{\bf k}_{1T}-{\bf k}_{2T})^2)}\,,
\end{equation}
where the transverse momentum dependence now sets the  factorization scale. 
Then the modified factorization formula in the transverse  {\it impact 
parameter} representation is given by
\begin{eqnarray}
F^{\rm hard}_M(Q^2)\!\!&=&\!\!\int^1_0
dxdy\int\frac{d^2b_1}{(2\pi)^2}\frac{d^2b_2}{(2\pi)^2}\nonumber\\
\times{\mathcal P}_{2;M}(x,b_1,P,\mu)\!\!\!\!&\,&\!\!\!\!\!
\tilde{T}_H(x,y,Q,b_1,b_2,\mu)\,\,{\mathcal P}_{2;M}(y,b_2,P',\mu)\,,
\end{eqnarray}
where the modified DA ${\mathcal P}_{2;M}(x_i,b_i,P_i,\mu)$ absorbs the large 
infrared logarithms into the Sudakov exponent $S_i$ \cite{St} (including also 
the evolution of the DA from the factorization scale $1/b_i$ to the scale 
$\mu$):
\begin{eqnarray}
{\mathcal P}_{2;M}(x_i,b_i,P_i\simeq Q,\mu)\!&=&\!{\rm exp}\left[-S_i(XQ,b_i,\mu)\right]\tilde{{\mathcal P}}_{2;M}(x_i,b_i,1/b_i)\,;\nonumber\\
S_i(XQ,b_i,\mu)=s(x_iQ\!\!\!&,&\!\!\!1/b_i)+s((1-x_i)\,Q,
1/b_i)+2\int^{\mu}_{1/b_i}\frac{d\bar{\mu}}{\bar{\mu}}\gamma_q(\alpha_s
({\bar{\mu}}^2))\,,\nonumber\\
s(XQ,1/b_i)=\int^{XQ/\sqrt{2}}_{1/b_i}\!\!\!\!\!\!&\,&\!\!\!\!\!\!
\frac{d\mu}{\mu}\left[\ln\left(\frac{XQ}{\sqrt{2}\mu}\right){\mathcal A}
(\alpha_s(\mu^2))
  + {\mathcal B}(\alpha_s(\mu^2))\right]\,, 
\end{eqnarray}
where $1/b_1$, $1/b_2$ set the factorization scales in the transverse impact 
configuration. In the above equations, the quark anomalous dimension is given 
by $\gamma_q(\alpha_s)=-\alpha_s/\pi$ and the ``cusp'' anomalous dimensions 
${\mathcal A}$ and ${\mathcal B}$, to one-loop accuracy are given by 
\begin{eqnarray}
{\mathcal A}(\alpha_s(\mu^2))\!&=&\!{\mathcal C}_F\frac{\alpha_s(\mu^2)}{
\pi}+\left[\left(\frac{67}{27}-\frac{\pi^2}{9}\right)N_c-\frac{10}{27}N_f+\frac{8}{3}\beta_0\ln
\left(\frac{e^{\gamma_E}}{2}\right)\right]\left(\frac{\alpha_s(\mu^2)}{\pi}\right)^2\,,\nonumber\\
{\mathcal B}(\alpha_s(\mu^2))\!&=&\!\frac{2}{3}\frac{\alpha_s(\mu^2)}{\pi}
\ln\left(\frac{e^{2\gamma_E-1}}{2}\right)
\end{eqnarray}
where the $\overline{\rm{MS}}$ running coupling to two-loop accuracy in 
standard perturbation theory is given by
\begin{equation}
\label{eq:coupling}
\frac{\alpha_s(\mu^2)}{\pi}=\frac{1}{\beta_0\ln(\mu^2/\Lambda^2_{\rm QCD})}-
\frac{\beta_1\ln(\ln(\mu^2/\Lambda^2_{\rm
    QCD}))}{\beta^3_0\ln^2(\mu^2/\Lambda^2_
{\rm QCD})}
 \end{equation}
with $\beta_0\!=(11N_c-2N_f)/12=\!9/4$ and $\beta_1\!=(51N_c-19N_f)/24=\!4$
for $N_c\!=\!N_f\!=\!3$. Note that the above modified factorization calls for 
introducing a scale hierarchy $XQ>1/b_i>\Lambda_{\rm QCD}$ (where $X=x_i, 
(1-x_i)$, $x_1=x$ and $x_2=y$) to separate the distinct contributions 
from the perturbative and non-perturbative kinematic regions without the 
possibility of a ``double counting''. Note that there exist other schemes of 
defining the running coupling involving power corrections, restoring the 
explicit Landau singularity and the analyticity at $Q^2=0$ (see e.g., 
Refs.~\cite{Shirkov,analytic} and also Section~\ref{sec:numerics} for details.)

At low momentum transfers, the modified infrared free DAs are often 
approximated with constituent quark masses which are different from the actual 
masses of the current quarks and usually chosen close to the intrinsic 
transverse scale $\Lambda_{\rm{QCD}}$ of the hadron structure, i.e., between 
200-500 MeV. These quark masses which effectively parametrize the QCD vacuum 
effects are also used to suppress possible endpoint effects. Hence, we have
\begin{equation}
\tilde{{\mathcal P}}_{2;M}(x_i,b_i,1/b_i)\simeq\tilde{{\mathcal
    P}}_{2;M}(x_i,b_i,1/b_i,{\mathcal M}_q)
\end{equation}
which could be expressed in terms of the full momentum-space light-cone wave 
function $\Psi_{2;M}$ (which also includes the transverse momentum 
distribution of the constituent bound state partons):
\begin{equation}
\tilde{{\mathcal P}}_{2;M}(x_i,b_i,1/b_i,{\mathcal M}_q)=\int_{{\mathbf k}^2_{iT}\leq
  (1/b_i)^2}\frac{d^2{\mathbf k}_{iT}}{16\pi^3} \,\Psi_{2;M}(x_i,{\mathbf k}_{iT},1/b_i,{\mathcal M}_q)\,.
\end{equation}
To model the intrinsic transverse momentum dependence of the meson wave 
functions, we use the Brodsky-Huang-Lepage (BHL) gaussian prescription
\cite{crit5,BHL}:
\begin{equation}
\Psi_{2;M}(x_i,{\mathbf k}_{iT},1/b_i,{\mathcal M}_q)=\Phi_{2;M}(x_i,1/b_i)\,\Sigma(x_i,{\mathbf k}_{iT},{\mathcal M}_q)
\end{equation}
with
\begin{eqnarray}
\Phi_{2;M}(x_i,1/b_i)\!&=&\!A_{2;M}\,\phi_{2;M}(x_i,1/b_i)\,,\\
\Sigma(x_i,{\mathbf k}_{iT},{\mathcal
  M}_q)\!&=&\!\frac{16\pi^2\beta_{2;M}^2}{x_i(1-x_i)}\,{\rm{exp}}\left[-\frac{\beta^2_{2;M}}{x_i(1-x_i)}\left({\mathbf k}^2_{iT}+{\mathcal M}^2_q\right)\right]\,,
\end{eqnarray}
assuming equal masses of the two constituent quarks within the meson. The 
parameters $A_{2;M},\,\beta_{2;M}$ and ${\mathcal M}_q$ are fixed using 
phenomenological constraints. The above integration then yields the full 
modified wave function in the impact representation: 
\begin{eqnarray}
\tilde{{\mathcal P}}_{2;M}(x_i,b_i,1/b_i,{\mathcal
  M}_q)\!&=&\!A_{2;M}\,\phi_{2;M}(x_i,1/b_i)\,{\rm{exp}}\left[-\frac{\beta^2_{2;M}{\mathcal
  M}^2_q}{x_i(1-x_i)}\right]\nonumber \\ 
&&\times{\rm{exp}}\left[-\frac{b^2_ix_i(1-x_i)}{4\beta^2_{2;M}}\right]
\end{eqnarray}
Including the RG evolution equation for the hard kernel:
\begin{equation}
\tilde{T}_H(x,y,Q,b_1,b_2,\mu)={\rm exp}\left[-4\int^t_{\mu}
\frac{d\bar{\mu}}{\bar{\mu}}\gamma_q(\alpha_s({\bar{\mu}}^2))\right]
\tilde{T}_H(x,y,Q,b_1,b_2,t)\,,
\end{equation}
where
\begin{equation}
t={\rm max}(\sqrt{xy}\,Q,1/b_1,1/b_2)\,,
\end{equation}
one arrives at the ``double-b'' factorization formula for the meson form
factor at the twist-2 level \cite{pff3} :
\begin{eqnarray}
\label{eq:ff-hard_t2}
F^{(t=2)}_M(Q^2)\!&=&\!16\pi Q^2{\mathcal C}_F\int^1_0 xdxdy\int^{\infty}_{0} 
b_1 db_1 b_2 db_2\,\alpha_s(t)\,\tilde{{\mathcal P}}_{2;M}(x,b_1,1/b_1,{\mathcal M}_q)\nonumber \\
&&\times\tilde{{\mathcal P}}_{2;M}(y,b_i,1/b_2,{\mathcal
  M}_q)\,H(x,y,Q,b_1,b_2)\,{\rm exp}\left[-S(x,y,b_1,b_2,Q)\right]\nonumber\\
&&
\end{eqnarray}
with
\begin{eqnarray}
\label{eq:kernel}
H(x,y,Q,b_1,b_2)\!&=&\!K_0(\sqrt{xy}\,Qb_2)\left[\theta(b_1-b_2)K_0(
\sqrt{x}\,Qb_1)I_0(\sqrt{x}\,Qb_2)\right.\nonumber\\
&&+\left. \theta(b_2-b_1)K_0(\sqrt{x}\,Qb_2)I_0(\sqrt{x}\,Qb_1)\right]\,.
\end{eqnarray}
$K_0$ and $I_0$ are modified Bessel functions and the full Sudakov exponent is 
given by
\begin{equation}
\label{eq:Sudakov}
S(x,y,b_1,b_2,Q)=\sum^2_{i=1}\left[s(x_iQ,1/b_i)+s((1-x_i)\,Q,1/b_i)+2
\int^{t}_{1/b_i}\frac{d\bar{\mu}}{\bar{\mu}}\gamma_q(\alpha_s({\bar{\mu}}^2))
\right]\,.
\end{equation}
For completeness, the expression for the full Sudakov factor $s(XQ,1/b_i)$, up 
to next-to-leading logarithm accuracy is given in the 
Appendix~\ref{app:sudakov}. The expression slightly differs from the result 
given in Ref.~\cite{LiBD}, but numerically this difference is insignificant at 
our working accuracy. Note that this difference was first observed in 
Ref.~\cite{pff5}.

To include the sub-leading twist-3 corrections to the form factor, the 
hard scattering kernel gets slightly modified as compared to the twist-2 case
which turns out to be \cite{pff2}
\begin{equation}
T^{(t=3)}_H(x,y,Q^2,{\bf k}_{1T},{\bf k}_{2T},\mu^2)=\frac{64\pi\,
{\mathcal C}_F\,\alpha_s(\mu^2)\,x}{(xQ^2+{\bf
  k}_{1T}^2)(xy\,Q^2+({\bf k}_{1T}-{\bf k}_{2T})^2)}\,.
\end{equation}
Applying the momentum projection operator \cite{Beneke,Wei},
\begin{equation}
{\mathcal M}^{M}_{\alpha\beta}=i\left\{P\!\!\!\!/\gamma_5\Psi_{2,M}-\mu_M
\gamma_5\left(\Psi^p_{3;M}-i\sigma_{\mu\nu}n^\mu\bar{n}^\nu\frac{\Psi^{\sigma
\,'}_{3;M}}{6}+i\sigma_{\mu\nu}P^\mu\frac{\Psi^{\sigma}_{3;M}}{6}
\frac{\partial}{\partial {k_T}_\nu}\right)\right\}_{\alpha\beta}
\end{equation}
on the bilocal matrix element with quark flavors $f_1$ and $f_2$ 
($f_{1,2}=u,d,s$),
\begin{eqnarray}
\left<0\left|{\bar q}_{f_1}(z)\,q_{f_2}(-z)\right|M(P)\right>_{\alpha\beta}\!&=&\!i
\int^1_0 dx\int d^2 {\bf k}_T\, e^{i\xi(zp)}\\
&&\times\left\{P\!\!\!\!/\gamma_5\Psi_{2;M}-\mu_M\gamma_5\left(\Psi^p_{3;M}-\sigma_{\mu\nu}P^\mu z^\nu\frac{\Psi^{\sigma}_{3;M}}{6}\right)\right\}_{\alpha\beta}\,,\nonumber
\end{eqnarray}
where $\Psi^{\sigma\,'}_{3;M}(x,{\bf k}_T,1/b,{\mathcal
  M}_q)=\partial\Psi^{\sigma}_{3;M}(x,{\bf k}_T,1/b,{\mathcal M}_q)/\partial
x$, $n=(1,0,{\bf 0}_T)$ is the unit vector in the ``+'' direction, 
$\bar{n}=(0,1,{\bf 0}_T)$ is the unit vector in the ``-'' direction,
$\xi=2x-1$ and $\mu_M=m^2_M/(m_{q_{f_1}}+m_{q_{f_2}})$, one obtains the 
final formula for hard meson form factor up to twist-3 corrections given by 
\cite{pff7,pff9}
\begin{eqnarray}
\label{eq:ff-hard}
F^{\rm hard}_M(Q^2)\!&=&\!F^{(t=2)}_M(Q^2)+F^{(t=3)}_M(Q^2)\nonumber\\
&=&\!32\pi Q^2{\mathcal C}_F\int^1_0 dxdy
\int^{\infty}_{0} b_1db_1 b_2 db_2\,\alpha_s(t)\nonumber\\
&&\times\left[\frac{x}{2}\,\tilde{{\mathcal
      P}}_{2;M}(x,b_1,1/b_1,{\mathcal M}_q)\,\tilde{{\mathcal
      P}}_{2;M}(y,b_2,1/b_2,{\mathcal M}_q)\right.
\nonumber\\
&&+\left.\frac{\mu^2_M}{Q^2}\left({\bar x}\,\tilde{{\mathcal
      P}}^{\,p}_{3;M}(x,b_1,1/b_1,{\mathcal M}_q)\,\tilde{{\mathcal
      P}}^{\,p}_{3;M}(y,b_2,1/b_2,{\mathcal M}_q)\right.\right.\nonumber\\
&&+\left.\left.\frac{(1+x)}{6}\,\tilde{{\mathcal P}}^{\,p}_{3;M}
(x,b_1,1/b_1,{\mathcal M}_q)\,\tilde{{\mathcal P}}^{\,\sigma\,'}_{3;M}
(y,b_2,1/b_2,{\mathcal M}_q)\right.\right.\nonumber\\
&&+\left.\left.\frac{1}{2}\tilde{{\mathcal
        P}}^{\,p}_{3;M}(x,b_1,1/b_1,{\mathcal M}_q)\,\tilde{{\mathcal
        P}}^{\,\sigma}_{3;M}(y,b_2,1/b_2,{\mathcal M}_q)
\right)\right]\,H(x,y,Q,b_1,b_2)\nonumber\\
&&\times\prod^2_{i=1}S_t(x_i)\,S_t({\bar x_i})\,{\rm
  exp}\left[-S(x,y,b_1,b_2,Q)\right]\,\,;\,\,{\bar x_i}=1-x_i\,.
\end{eqnarray}
Here, we have assumed a similar gaussian ansatz in the transverse momentum
distribution of the modified twist-3 wave functions:
\begin{eqnarray}
\tilde{{\mathcal P}}^{\,p}_{3;M}(x_i,b_i,1/b_i,{\mathcal M}_q)
&=&\!A^{p}_{3;M}\,\phi^{\,p}_{3;M}(x_i,1/b_i)\,{\rm{exp}}\left[-\frac{(\beta^p_{3;M})^2{\mathcal
      M}^2_q}{x_i(1-x_i)}\right]\nonumber\\
&&\times{\rm{exp}}\left[-\frac{b^2_i\,x_i(1-x_i)}{4(\beta^p_{3;M})^2}\right] \, ,\nonumber
\end{eqnarray}
\begin{eqnarray}
\tilde{{\mathcal P}}^{\,\sigma}_{3;M}(x_i,b_i,1/b_i,{\mathcal M}_q)
&=&\!A^{\sigma}_{3;M}\,\phi^{\sigma}_{3;M}(x_i,1/b_i)\,{\rm{exp}}\left[-\frac{(\beta^{\sigma}_{3;M})^2{\mathcal
      M}^2_q}{x_i(1-x_i)}\right]\nonumber\\
&&\times{\rm{exp}}\left[-\frac{b^2_i\,x_i(1-x_i)}{4(\beta^{\sigma}_{3;M})^2}\right]\,.
\end{eqnarray}
The hard kernel $H$ and the Sudakov exponent $S$ are given by 
Eq.~(\ref{eq:kernel}) and Eq.~(\ref{eq:Sudakov}), respectively. The above 
formula is used to evaluate the pion and kaon hard form factors using the 
twist-2 DAs \,Eq.~(\ref{eq:pion-DA2}) and Eq.~(\ref{eq:kaon-DA2}), 
respectively, and twist-3 DAs provided in Appendix~\ref{app:DA}. The $S_t(x_i)$
are {\it jet functions}, defined as eikonalized matrix elements of quark fields
attached by a Wilson line, arising from another kinematic resummation scheme 
called the {\it threshold resummation}, as introduced in Ref.~\cite{Sanda,Li}. 
The modified treatment of the collinear factorization prescription works 
reasonably well for the twist-2 case but for the twist-3 case, the Sudakov 
suppression factor may still not be effective enough in shielding the 
non-perturbative enhancements due to endpoint singularities. These are 
kinematic singularities of the scattering amplitude when the longitudinal 
momentum fraction $x$ of the valence partons (quarks) go to 0, 1. Therefore, in
addition there is a need to sum up the collinear double logarithms of the type 
$\alpha_s\ln^2 x$ to all orders, which are then collected into these jet 
functions. The  exact form of $S_t(x_i)$ involves a one parameter integration, 
but for the sake of numerical calculations it is convenient to take the simple 
parameterization, as proposed in Ref.~\cite{Sanda,Li} :
\begin{equation}
 S_t(x_i)=\frac{2^{1+2c}\Gamma(3/2+c)}{\sqrt{\pi}\Gamma(1+c)}\left[x_i(1-x_i)
\right]^c\,,
\end{equation}
where the parameter $c\approx 0.3$ for light pseudo-scalar mesons like the pion
and kaon. The jet functions vanish at the endpoints and modify the 
endpoint behavior of the DAs, providing enough suppression to damp the 
artificial effect of endpoint singularities.
  
\subsection{Soft contributions via Local Duality}
The perturbative predictions for the pion form factor are known to be
relatively small for phenomenological low momentum transfers ($Q^2\leq 10$ 
GeV$^2$) \cite{pff01,pff02,pff03,crit4,crit1,crit7,crit9}, as also evident from
our analysis in the next section. Clearly, there is the need for including 
non-factorizable soft contributions to explain the experimental data. The 
factorization ansatz Eq.~(\ref{eq:factor}) holds for large momentum transfers 
under the assumption that only the contributions from valence parton states 
dominate. This approximation no longer holds true at small momenta when 
contributions from higher Fock states with more than valence partons become 
significant. In addition, there could be non-perturbative enhancements from the
 so-called {\it Feynman mechanism}, which corresponds to selecting a hadronic
configuration in which one of the valence parton carries almost the entire 
hadron momenta. Unfortunately, due to the complexity of soft QCD processes, 
there are no unambiguous ways to calculate these contributions analytically 
using the parton picture and Feynman diagrammatics, other than using 
theoretical models for the DAs. In this paper, we follow the Local Duality 
(LD) approach from QCD sum rules as in Ref.~\cite{pff6} where the same 
problem is addressed without a direct reference to DAs. In this section, we 
simply use the result for the soft form factor derived in the LD approach:
\begin{equation}
\label{eq:ff-soft}
F^{\rm soft}_M(Q^2)=F^{\rm LD}_M(Q^2)=1-\frac{1+6s_0(Q^2)/Q^2}
{(1+4s_0(Q^2)/Q^2)^{3/2}}\,.
\end{equation} 
The {\it duality interval} $s_0$, encodes the non-perturbative information 
about higher excited states and continuum contributions and is given by
\begin{equation}
s_0(Q^2)=4\pi^2f^2_M/(1+\frac{\alpha_s(Q^2)}{\pi})\,.
\end{equation} 
Expanding in inverse powers of $Q$ gives $F^{\rm soft}_M(Q^2)\sim 1/Q^4$ for 
large $Q^2$ and is thus expected to be sub-leading compared to the leading 
perturbative contribution from Eq.~(\ref{eq:hard-asy}). Nevertheless, at low 
and moderate momentum transfers the soft contributions turn out to be so 
significant in obtaining a good agreement with the experimental data. This
fact is clearly revealed in our analysis in the next section. 

Next we add together the hard and the soft contribution to obtain 
the total contributions to the electromagnetic form factor $F_M(Q^2)$. Here,
it is necessary to ensure that the respective contributions lie within their 
domains of validity to minimize the possibility of a double counting. This  
technique, as introduced in Ref.~\cite{pff6}, employs gauge invariance that 
protects the value $F_M(0)=1$, through the vector Ward identity relating a 
3-point Green function to a 2-point Green function at zero momentum transfer,
i.e., $F^{\rm LD}_M(Q^2\!=\!0)\!=\!1$. This implies that $F^{\rm
  hard}_M(Q^2\!=\!0)\!=\!0$. A ``smooth'' transition from the hard to the soft
behavior is then ensured by a {\it matching ansatz}  from the large $Q^2$
behavior (arising from $F^{\rm hard}_M(Q^2)$) to the low $Q^2$ behavior
(arising from $F^{\rm soft}_M(Q^2)$). This can be done by introducing a mass 
scale $M_0$ which in the LD approach should be identified with the threshold 
$M^2_0=2s_0$. The twist-2 part of the hard form factor $F^{(t=2)}_M(Q^2)$ 
is then modified following Ref.~\cite{pff6}
\begin{equation}
F^{(t=2)}_M(Q^2)\rightarrow \left(\frac{Q^2}{2s_0(Q^2)+Q^2}\right)^2 F^{(t=2)}_M(Q^2)\,.
\end{equation}
However, for the twist-3 case, the ``matching function'' $\Phi(z)=1/(1+z)^2$,
with $z=Q^2/M_0^2$ is insufficient to ensure the Ward indentity at $Q^2=0$. 
To correct for the singular ($\sim 1/Q^4$) behavior, we make a similar 
modification of the twist-3 part via the replacement
\begin{equation}
F^{(t=3)}_M(Q^2)=\tilde{F}^{(t=3)}_M(Q^2)\frac{M^4_0}{Q^4}\rightarrow\tilde{F}^{(t=3)}_M(Q^2)\frac{M^4_0}{M^4_0+Q^4}
\end{equation}
with the choice of the matching function $\tilde{\Phi}(z)=1/{(1+z^2)}^2$. 
This yields the Ward identity corrected 
twist-3 part:
\begin{equation}
F^{(t=3)}_M(Q^2)\rightarrow \left(\frac{Q^4}{4s^2_0(Q^2)+Q^4}\right)^2 F^{(t=3)}_M(Q^2)\,.
\end{equation}
Finally, we arrive at our expression for the total electromagnetic form factor 
for a charged meson $M\,(\pi^\pm,K^\pm$), valid for all values of $Q^2$ and is 
given by
\begin{eqnarray}
\label{eq:total_ff}
F_M(Q^2)\!&=\!&1-\frac{1+6s_0(Q^2)/Q^2}{(1+4s_0(Q^2)/Q^2)^{3/2}}\\
&&+\left(\frac{Q^2}{2s_0(Q^2)+Q^2}\right)^2 F^{(t=2)}_M(Q^2)+\left(\frac{Q^4}{4s^2_0(Q^2)+Q^4}\right)^2 F^{(t=3)}_M(Q^2)\nonumber\,,
\end{eqnarray}
where $F^{(t=2)}_M(Q^2)$ and $F^{(t=3)}_M(Q^2)$ are given from 
Eq.~(\ref{eq:ff-hard_t2}) and Eq.~(\ref{eq:ff-hard}), respectively.

\section{Numerical Results}
\label{sec:numerics}
\begin{figure}
    \begin{tabular}{cc}
      \hspace{-1.4cm}\resizebox{9cm}{!}{\includegraphics{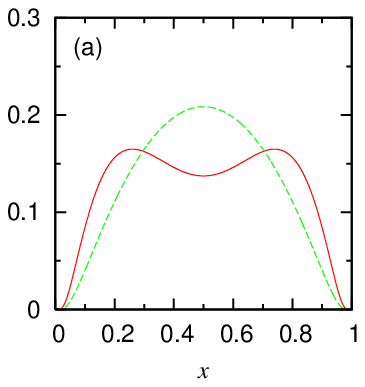}} &
      \resizebox{9cm}{!}{\includegraphics{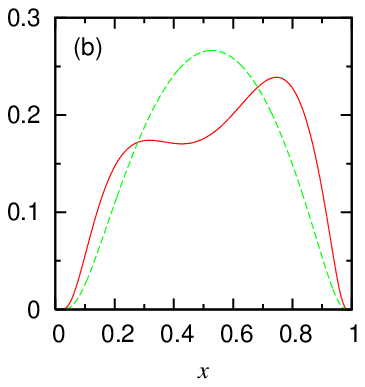}} \\
    \end{tabular}
    \caption{Twist-2 light-cone wave functions for (a) the pion  
      $\tilde{\mathcal P}_{2;\pi}$ and (b) the kaon  
      $\tilde{\mathcal P}_{2;K}$  (solid lines), along with the wave 
      functions corresponding to the respective asymptotic DAs (dashed
      lines). The DAs are defined at the scale $\mu_0=1$ GeV.}
    \label{fig:piDAs}
\end{figure}

\begin{figure}
    \hspace{-1.5cm}\begin{tabular}{cc}
      \resizebox{8.8cm}{!}{\includegraphics{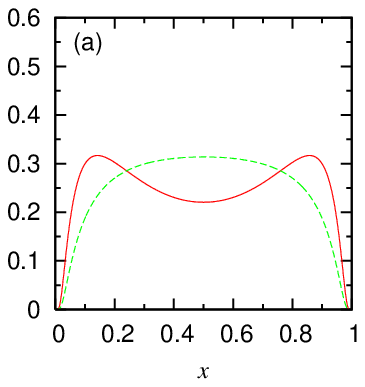}} &
      \resizebox{8.8cm}{!}{\includegraphics{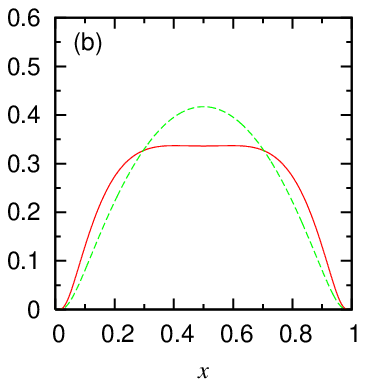}} \\
      \resizebox{8.8cm}{!}{\includegraphics{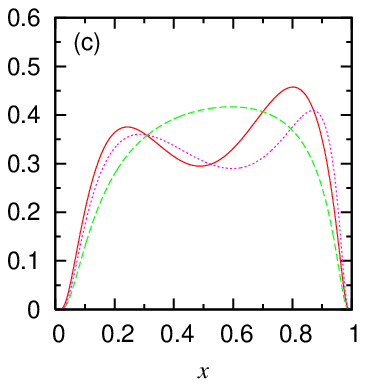}} &
      \resizebox{8.8cm}{!}{\includegraphics{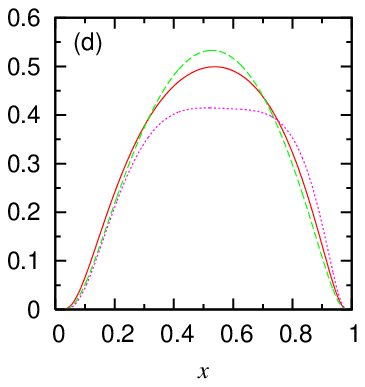}} \\
    \end{tabular}
    \caption{2-particle twist-3 light-cone wave functions for the pion (a) 
      $\tilde{\mathcal P}^{p}_{3;\pi}$ \& (b) 
      $\tilde{\mathcal P}^{\sigma}_{3;\pi}$ and for the kaon (c) 
      $\tilde{\mathcal P}^{p}_{3;K}$ \& (d) $\tilde{\mathcal
      P}^{\sigma}_{3;K}$  with G-parity even terms (solid lines), along with 
      the wave functions corresponding to the respective asymptotic DAs (long 
      dashed lines). For the kaon, the twist-3 wave functions including 
      G-parity odd terms are also shown (dotted lines). The DAs are defined at 
      the scale $\mu_0=1$ GeV.}
    \label{fig:KDAs}
\end{figure}

At first, we need to determine the pion and kaon gaussian parameters
$A_{2;M}$, $A^p_{3;M}$, $A^{\sigma}_{3;M}$ and 
$\beta_{2;M},\,\beta^p_{3;M},\,\beta^{\sigma}_{3;M}$ for the twist-2 and 
twist-3 light-cone wave functions, respectively. For the pion, they are 
obtained from the two constraints: firstly, by virtue of the leptonic decay 
$\pi\rightarrow \mu \nu_\mu$, we have the condition
\begin{equation}
\int^1_0 dx\,\int \frac{d^2{\mathbf k}_{T}}{16\pi^3}\,\Psi_{\pi}(x,{\mathbf k}_T,{\mathcal M}_{u,d})=\frac{f_\pi}{2\sqrt{2N_c}}\,,
\end{equation}
leading to
\begin{equation}
A_\pi\int^1_0 dx\,\phi_{\pi}(x)\,{\rm exp}\left[-\frac{\beta^2_\pi\,{\mathcal M}^2_{u,d}}{x(1-x)}\right]=\frac{f_\pi}{2\sqrt{6}}
\end{equation}
and secondly, from $\pi^0\rightarrow\gamma\gamma$, we have the condition
\begin{equation}
\label{eq:pi-gamma-gamma}
\int^1_0 dx\,\Psi_{\pi}(x,{\mathbf k}_T\!=\!0,{\mathcal M}_{u,d})=\frac{\sqrt{2N_c}}{f_\pi}\,,
\end{equation}
which implies
\begin{equation}
16A_\pi\beta^2_\pi\pi^2\int^1_0 dx\,\frac{\phi_\pi(x)}{x(1-x)}\,{\rm exp}\left[-\frac{\beta^2_\pi\,{\mathcal M}^2_{u,d}}{x(1-x)}\right]=\frac{\sqrt{6}}{f_\pi}\,,
\end{equation}
where we use the constituent quark mass $\mathcal M_{u,d}=0.33$ GeV for both 
the $u$ and $d$ valence quarks in the pion. In the case of the kaon, firstly from
the leptonic decay $K\rightarrow\mu \nu_{\mu}$, we have the 
constraint
\begin{equation}
A_K\int^1_0 dx\,\phi_{K}(x)\,{\rm exp}\left[-\beta^2_K\left(\frac{{\mathcal
        M}^2_{s}}{x}+\frac{{\mathcal
        M}^2_{u,d}}{1-x}\right)\right]=\frac{f_K}{2\sqrt{6}}\,\,.
\end{equation}
As for the second constraint, no straightforward condition like 
Eq.~(\ref{eq:pi-gamma-gamma}) could be obtained for the kaon. On the other 
hand, by virtue of SU(3) isospin symmetry, it is reasonable to make an 
assumption that for the kaon the average transverse momentum-squared of the 
valence partons defined by
\begin{equation}
\left<{\mathbf k}^2_T\right>_K=\frac{\int dx \int d^2{\mathbf k}_{T}
  \,\left|{\mathbf k}^2_{T}\right| \left|\Psi_{K}(x,{\mathbf k}_{T},{\mathcal
    M}_{u,d,s})\right|^2}{\int dx \int d^2{\mathbf k}_{T} \,\left|\Psi_{K}(x,{\mathbf k}_{T},{\mathcal M}_{u,d,s})\right|^2}\,,
\end{equation}
has about the same value as in the case of the pion. We have checked that for both the 
twist-2 and twist-3 pion wave functions $\left<{\mathbf
    k}^2_T\right>^{1/2}_\pi\approx0.35$ GeV. This yields our second condition 
for determining the wave function parameters:
\begin{equation}
(0.35)^2\approx\frac{1}{2\beta^2_K}\,\frac{\int^1_0 dx\,\phi^2_K(x)\,{\rm exp}\left[-2\beta^2_K\left(\frac{{\mathcal M}^2_{s}}{x}+\frac{{\mathcal M}^2_{u,d}}{1-x}\right)\right]}{\int^1_0 dx\,\frac{\phi^2_K(x)}{x(1-x)}\,{\rm exp}\left[-2\beta^2_K\left(\frac{{\mathcal M}^2_{s}}{x}+\frac{{\mathcal M}^2_{u,d}}{1-x}\right)\right]}\,,
\end{equation}
where ${\mathcal M}_s=0.45$ GeV is used as the constituent $s$-quark mass and 
the full light-cone kaon wave function is given by
\begin{equation}
\Psi_K(x,{\mathbf k}_{T},{\mathcal M}_{u,d,s})=\frac{16\pi^2\beta^2_KA_K}{x(1-x)}\,\phi_K(x)\,{\rm{exp}}\left[-\beta^2_K\left(\frac{{\mathbf k}^2_{T}+{\mathcal M}^2_{s}}{x}+\frac{{\mathbf k}^2_{T}+{\mathcal M}^2_{u,d}}{1-x}\right)\right]
\end{equation}
with $x$ being the longitudinal momentum fraction of the $s$ quark. For our 
numerical analysis we use typical ``double-humped'' type
\cite{Brodsky,Efremov,Radyushkin} DAs $\phi_{\pi,K}(x,\mu^2)$, derived in the 
framework of QCD sum rules \cite{DAs1,DAs2,DAs4}. Note that we have considered 
$N_c=3$ in the expressions for the DAs. In Fig.~\ref{fig:piDAs} and 
Fig.~\ref{fig:KDAs}, we display the twist-2 and twist-3 light-cone wave 
functions for the pion and kaon, respectively, along with their corresponding 
asymptotic wave functions. Note that the plots exclude the normalization 
factors of $\frac{1}{2\sqrt{6}}$ and $\frac{1}{4\sqrt{6}}$ for the individual 
DAs to facilitate comparison with one another. All the DAs are defined at the 
scale $\mu_0\!=\!1$ GeV. The twist-2 and twist-3 DA input parameters are taken 
from Table 3 of Ref.~\cite{DAs4} which we again provide in 
Table~\ref{table:para} along with the rest of the input parameters for the 
wave functions. Note that for the kaon we have shown both the type of wave 
functions, i.e., with and without including the G-parity-breaking terms.

   
\begin{table}[h]
\begin{center} 
\begin{tabular}{|c|c||c|c||c|}
\hline
$\pi^\pm$ & At $\mu_0=1$ GeV & $K^\pm $ & At $\mu_0=1$ GeV & units\\
\hline\hline
- & - & $m_{u,d}$ & $5.6\pm 1.6$ \cite{DAs4} & MeV \\
- & - & $m_s$ & $137\pm 27$ \cite{DAs4} & MeV \\
${\mathcal M}_{u,d}$ & $0.33$ & ${\mathcal M}_{u,d}$ & $0.33$ & GeV  \\
- & - & ${\mathcal M}_{s}$ & $0.45$ & GeV  \\
$m_\pi$         & $139$ & $m_K$         & $493$ & MeV  \\
$a^\pi_1$ & 0 & $a^K_1$       & $0.06\pm 0.03$ \cite{DAs4} & - \\ 
$a^\pi_2$ & $0.25\pm 0.15$ \cite{DAs4}& $a^K_2$ & $0.25\pm 0.15$ \cite{DAs4} &
- \\
$f_{\pi}$ & $131$  & $f_{K}$ & $1.22\,|f_\pi|$ \cite{wfs0} & MeV\\
$f_{3\pi}$ & $0.0045\pm 0.0015$ \cite{DAs4}& $f_{3K}$ & $0.0045\pm
0.0015$ \cite{DAs4} & GeV$^2$ \\ 
$\omega_{3\pi}$ & $-1.5\pm 0.7$ \cite{DAs4}& $\omega_{3K}$ & $-1.2\pm 0.7$
\cite{DAs4} & -\\ 
$\lambda_{3\pi}$& 0 & $\lambda_{3K}$& $1.6\pm 0.4$ \cite{DAs4} & - \\ 
\hline
\end{tabular}
\end{center}
\caption{Various input hadronic parameters for twist-2 and twist-3 light-cone 
wave functions at $\mu_0=1$ GeV.}
\label{table:para}
\end{table}

\begin{table}[h]
\begin{center} 
\begin{tabular}{|c|c||c|c|c||c|}
\hline
$\pi^\pm$ & G (even) & $K^\pm $ & G (even) &  G (even + odd) & units \\
\hline\hline
$A_{2;\pi}$ & $1.69\,(1.66)_{\rm as}$ & $A_{2;K}$ & $2.06\,(2.07)_{\rm as}$ &
$2.06\,(2.07)_{\rm as}$ & - \\ 
$A^{p}_{3;\pi}$ & $3.76\,(3.59)_{\rm as}$ & $A^{p}_{3;K}$ & $4.40\,(4.56)_{\rm
  as}$ &
$4.35\,(4.56)_{\rm as}$ & - \\ 
$A^{\sigma}_{3;\pi}$ & $3.37\,(3.33)_{\rm as}$ & $A^{\sigma}_{3;K}$ & 
$4.16\,(4.14)_{\rm as}$ & $4.06\,(4.14)_{\rm as}$ & - \\ 
$(\beta_{2;\pi})^2$ & $0.76\,(0.87)_{\rm as}$ & $(\beta_{2;K})^2$ & 
$0.78\,(0.89)_{\rm as}$ & $0.78\,(0.89)_{\rm as}$ & GeV$^{-2}$ \\ 
$(\beta^{p}_{3;\pi})^2$ & $0.62\,(0.74)_{\rm as}$ & 
$(\beta^{p}_{3;K})^2$ & $0.70\,(0.79)_{\rm as}$ & $0.65\,(0.79)_{\rm
  as}$ & GeV$^{-2}$ \\ 
$(\beta^{\sigma}_{3;\pi})^2$ & $0.81\,(0.87)_{\rm as}$ & 
$(\beta^{\sigma}_{3;K})^2$ & $0.88\,(0.89)_{\rm as}$ &
$0.84\,(0.89)_{\rm as}$ & GeV$^{-2}$ \\ \hline\hline
$\chi^{\rm fit}_{3\pi}$& $1.3\pm 0.4$ & $\chi^{\rm fit}_{3K}$ & $1.3
\pm 0.4$ & $1.3\pm 0.4$ & GeV \\
\hline
\end{tabular}
\end{center}
\caption{Various determined hadronic parameters for twist-2 and twist-3
  light-cone wave functions at $\mu_0=1$ GeV. The numbers in the parentheses
  correspond to values for the asymptotic wave functions.}
\label{table:para-det} 
\end{table}
In Refs.~\cite{DAs1,DAs2,DAs4}, the DAs were assumed to obey equations of 
motion (EOM) of on-shell quarks for which 
$\mu_{M}\!=\!m^2_{M}/(m_{q}+m_{q,s})\!\approx\!1.7$ GeV was used. This is 
not strictly correct, since the quarks are not exactly on-shell but instead 
confined within the hadrons. We therefore prefer using a ``chiral-enhancement''
parameter $\chi_{3M}(1 {\rm GeV})\!\approx\!\mu_\pi\!\approx\!\mu_K$, instead 
of $\mu_M$ in both the DAs and also the expression for the hard form factor 
Eq.~(\ref{eq:ff-hard}). Its numerical value is fixed by fitting the total 
form factor Eq.~(\ref{eq:total_ff}) to the available ``world-data'' for the 
pion \cite{pion_exp1,pion_exp2,pion_exp3,pion_exp4,pion_exp5,pion_exp6,pion_exp7,pion_exp8}. Note that in this fitting procedure only the asymptotic 
forms of the twist-2 and twist-3 DAs (Eq.~(\ref{eq:asy}) \& 
Eq.~(\ref{eq:twist-3asy})) were used in the pion wave functions. The running 
behavior $\chi_{3M}(\mu)$ is then later introduced while calculating the form 
factors whose RG behavior is assumed to be the same as that of $\mu_{M}$ (see, 
Appendix \ref{app:DA}). In other words, this amounts to the replacement 
$\mu^2_M\rightarrow \chi_{3M}(1/b_1)\,\chi_{3M}(1/b_2)$ in 
Eq.~(\ref{eq:ff-hard}). Also, as the bulk of the world pion data is 
concentrated in the very low energy region where the usual running coupling 
rapidly diverges, we in addition use an analytic prescription for the QCD 
running coupling to prove our results. The analytic scheme was suggested 
originally in Ref.~\cite{analytic} for calculating the pion form factor and 
further developed in Refs.~\cite{pff5,pff6} for NLO calculations. Here, the 
central idea is the removal of the explicit Landau singularity present in 
perturbation theory rendering the coupling constant IR stable and reducing the 
IR sensitivity of perturbatively calculated hadronic observables. The scheme is
also known to display higher loop stability. Now, the usual two-loop running
coupling Eq.~(\ref{eq:coupling}) in standard pQCD can be approximately 
expressed via the {\it Lambert}\, $W_{-1}$ function
\begin{equation}
\frac{\alpha_s(\mu^2)}{\pi}=-\frac{\beta_0}{\beta_1}\left[1+W_{-1}\left
(-\frac{\beta^2_0}{\beta_1 e}\left(\frac{\Lambda^2_{\rm QCD}}{\mu^2}\right)\right)\right]^{-1}\,.
\end{equation}
The extension of the above formula in analytic perturbation theory is too 
complicated to be evaluated exactly and instead there is an alternate 
approximate expression in the $\overline{\rm{MS}}$ scheme, as suggested in 
Ref.~\cite{Shirkov} :
\begin{equation}
\label{eq:coupling_ana}
\frac{\alpha^{\rm
    an,approx}_s(\mu^2)}{\pi}=\frac{1}{\beta_0}\left[\frac{1}{l}+\frac{1}{1-{\rm{exp}}(l)}\right]\,;
\end{equation}
\begin{equation}
l=\ln\left(\frac{\mu^2}{\Lambda^2_{\rm
      an}}\right)+\frac{\beta_1}{\beta^2_0}\ln\sqrt{\ln^2\left
(\frac{\mu^2}{\Lambda^2_{\rm an}}\right)+4\pi^2}\,,
\end{equation}
where $\Lambda_{\rm an}$ in the analytic scheme is the analog of 
$\Lambda_{\rm{QCD}}$ in the usual perturbation theory and is chosen to be 
around 0.4 GeV for $N_f=3$. We use this formula for the analytic 
coupling in our calculations. The simple one-parameter fitting of the total 
form factor to the experimental data gives the best fit values of 1.2 GeV and
1.4 GeV for the usual and analytic QCD coupling schemes, respectively. Here,
we choose the average value $\chi_{3M}\!=\!1.3$ GeV for both the schemes, 
and generously consider the resulting difference from the phenomenological 
value of 1.7 GeV to contribute to the theoretical error, i.e., $\pm 0.4$ GeV. 
Note that the fitting takes into account the individual error-bars of the data 
points.

Finally, following Ref.~\cite{Li} the Sudakov suppression factor 
${\rm exp}(-s(XQ,1/b))$ is set to unity for small transverse separation 
``$b$'' between the valence quarks, i.e., whenever $b\!\!<\!\!\sqrt{2}/(XQ)$.
Also, to avoid probing into certain kinematic regions where ${\rm exp}(-S) $ 
may become greater than unity causing an enhancement instead of a suppression, 
${\rm exp}\,(-S)$ is set to 1 for 
$S<0$.

\subsection{The pion form factor}
\label{sec:pion}
\begin{figure}[t]
    \hspace{-2.5cm}\begin{tabular}{cc}
      \resizebox{9.1cm}{!}{\includegraphics{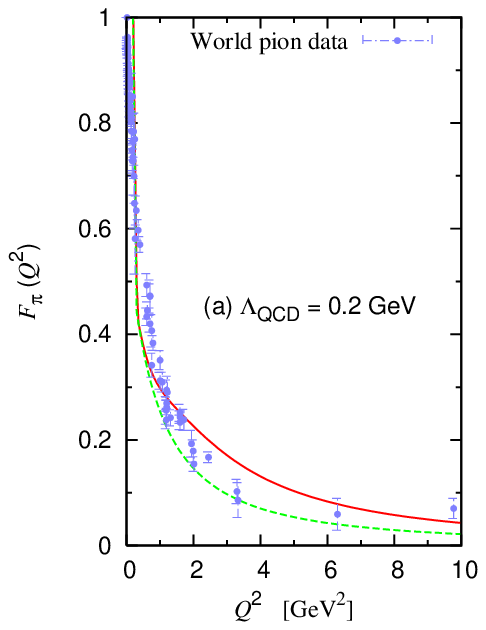}} &
      \resizebox{9.1cm}{!}{\includegraphics{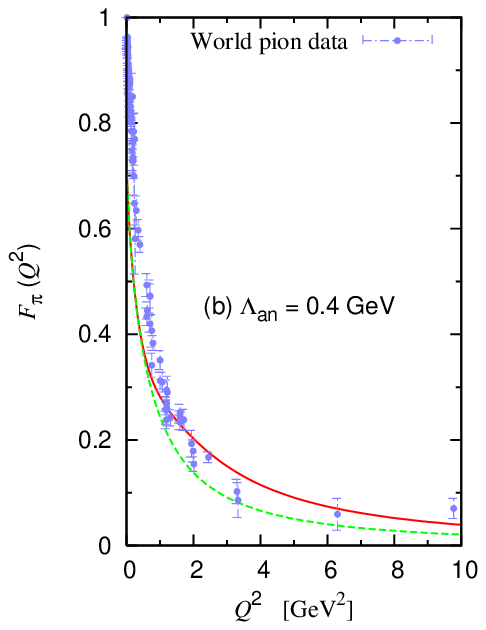}} \\
    \end{tabular}
    \caption{The total electromagnetic pion form factor at the twist-2 level
      (Soft+Twist-2), denoted by the dashed lines and at the twist-3 level
      (Soft+Twist-2+Twist-3), denoted by the solid lines, with (a) the usual 
      QCD running coupling and (b) the analytical QCD running coupling. The 
      world pion data are taken from Refs.~\cite{pion_exp1,pion_exp2,pion_exp3,pion_exp4,pion_exp5,pion_exp6,pion_exp7,pion_exp8}.}
    \label{fig:pi_ff}
\end{figure}

Using the DAs in Appendix A, we evaluated the total electromagnetic form
factor for the pion Eq.~(\ref{eq:total_ff}) using both the usual two-loop QCD 
running coupling Eq.~(\ref{eq:coupling}) and the analytical prescription 
Eq.~(\ref{eq:coupling_ana}). Fig.~\ref{fig:pi_ff} shows our results for the 
total form factor, along with the experimental ``world-data'' \cite{pion_exp1,pion_exp2,pion_exp3,pion_exp4,pion_exp5,pion_exp6,pion_exp7,pion_exp8} for the
pion. The plots correspond to $\Lambda_{\rm QCD}=0.2$ GeV and 
$\Lambda_{\rm an}=0.4$ GeV, respectively. It appears that the full twist-3 
calculations improve the agreement with experimental data to a much 
better extent at intermediate energies down to around 1-2 GeV$^2$ than for the
twist-2 case. Note that in the usual perturbative scheme, as $Q^2\rightarrow 0$
the total form factor becomes very unpredictable and starts oscillating 
between large values although $F_{\pi}(Q^2=0)=1$, satisfying the Ward 
identity. This clearly signals the breakdown of perturbation theory at such
small momentum transfers. 

\begin{figure}
    \hspace{-1.7cm}\begin{tabular}{cc}
      \resizebox{8.8cm}{!}{\includegraphics{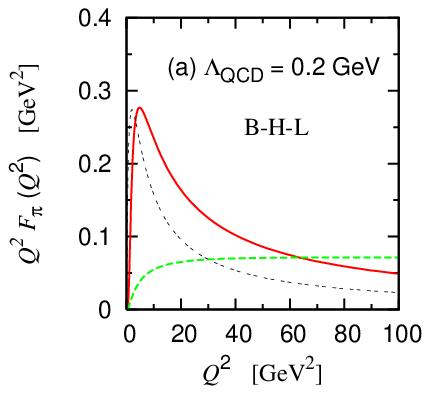}} &
      \resizebox{8.8cm}{!}{\includegraphics{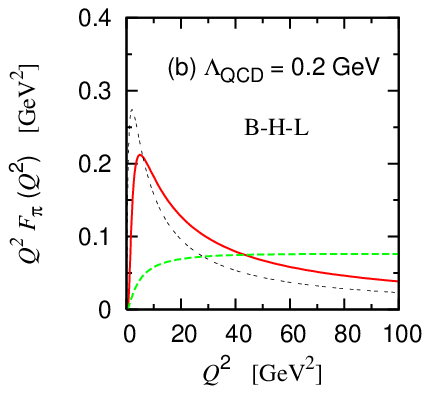}} \\
      \resizebox{8.8cm}{!}{\includegraphics{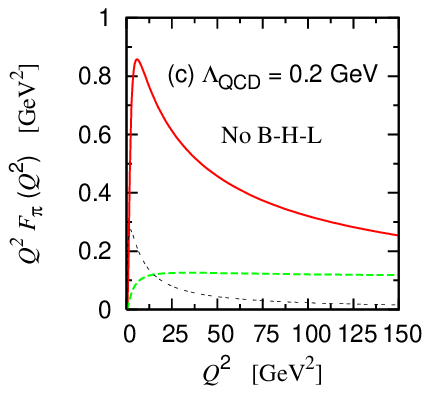}} &
      \resizebox{8.8cm}{!}{\includegraphics{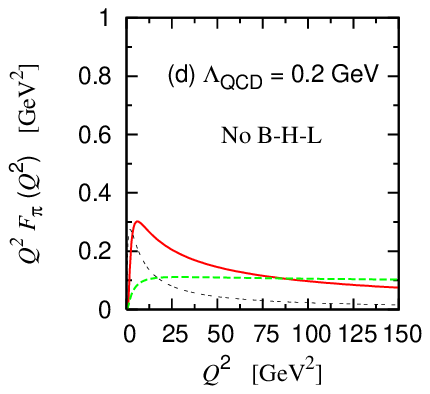}} \\
    \end{tabular}
    \caption{Twist-2 (long dashed lines) and Twist-3 (solid lines) corrections
      to the scaled hard pion form factor with the usual QCD running coupling. 
      Plots (a) \& (b) are obtained using BHL ansatz, while (c) \& (d) are 
      obtained with $\tilde{\mathcal P}_{M}(x)\rightarrow\phi_{M}(x)$. Also, 
      the plots (a) \& (c) do not include threshold resummation in the hard 
      contributions, which are included in (b) \& (d). The soft
      corrections (short dashed lines) are also shown.}
    \label{fig:pi_ff_hard}
\end{figure}

To study the contributions of endpoint effects and to distinguish 
individual soft and hard contributions, it is more useful to study the 
variation of the scaled pion form factor $Q^2 F_\pi$ with $Q^2$. In 
Fig.~\ref{fig:pi_ff_hard}, we show the individual contributions of the twist-2 
and twist-3 power correction to the scaled hard pion form factor over a wide 
range of momentum transfers for the usual QCD coupling. Clearly, the twist-3 
contributions are seen to be significantly larger than the leading twist-2 
counterparts at low momentum transfers, supporting the claims made in 
\cite{wfs1,wfs3,wfs4,pff2,pff7,pff9}. In fact, it is interesting to see the
endpoint enhancement in the twist-3 amplitudes much more explicitly if one 
rather considered only the collinear DAs to calculate the form factors in the 
usual perturbation theory, without considering the full transverse momentum 
dependence (e.g., the BHL ansatz) in the meson wave functions, as originally 
done in Ref.~\cite{Li}. In other words, one simply makes the replacement
$\tilde{\mathcal P}_{M}(x)\rightarrow\phi_{M}(x)$ in calculating the hard 
form factor. The inclusion of the transverse momenta and constituent quark 
masses in the wave function provides a natural cut-off for the soft and 
endpoint enhancements. Similar behavior can also be observed in the analytic 
case, although we have not displayed the enhancement which is less drastic. 
These facts suggest that the modified collinear factorization scheme, including
explicit transverse degrees of freedom with Sudakov suppression, which works 
well for the twist-2 case is not very effective at the twist-3 level in 
shielding such artificial non-perturbative enhancements at low momenta. To 
improve this situation, especially for the results obtained in usual 
perturbative  scheme, we use threshold resummation which along with Sudakov 
suppression provide large damping of the endpoint effects in the twist-3 
amplitude. The twist-2 part on the other hand remains mostly unaltered, if not 
slightly enhanced due to the threshold resummation, especially in the 
low-energy region. Note that in this respect the use of threshold resummation 
in the analytic scheme is somewhat redundant and has little effect on both the 
twist corrections. Finally, as expected, one observes that the twist-3 
corrections fall off rapidly with increasing $Q^2$ and beyond a certain point 
fall below the twist-2 corrections. At asymptotically large momentum transfers,
only the twist-2 contributions are expected to dominate.

\begin{figure}
      \begin{center}
      \resizebox{13.8cm}{!}{\includegraphics{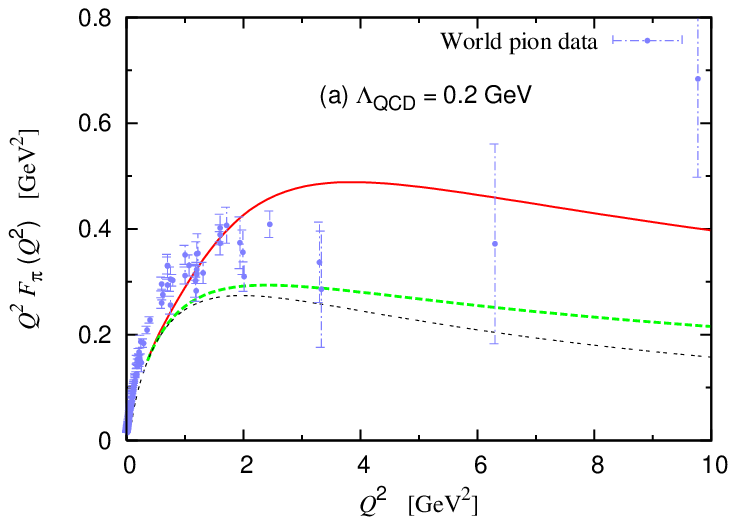}}\\
      \resizebox{13.8cm}{!}{\includegraphics{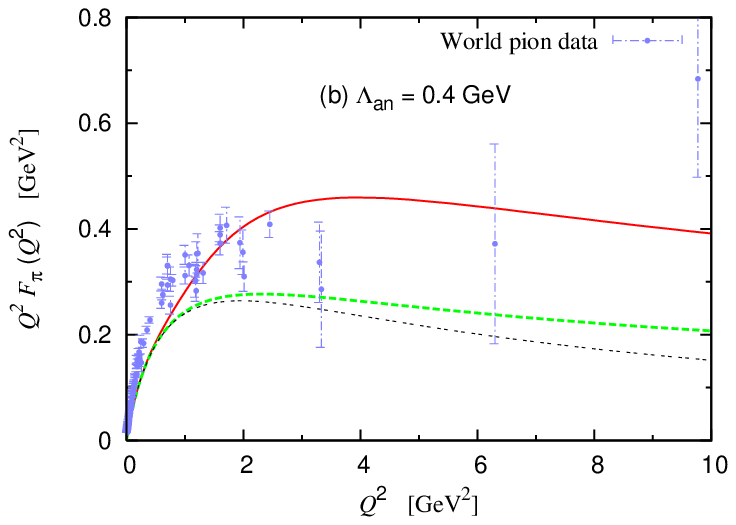}}
      \end{center}
      \caption{Scaled total electromagnetic form factor for the pion with (a)
          the usual QCD running coupling and (b) the analytical running
          coupling in the low and intermediate energy regime. The solid line
          represents the full twist-3 result (Soft+Twist-2+Twist-3), the
          long dashed lines represent the twist-2 result (Soft+Twist-2) 
          and the soft corrections are indicated by the short dashed lines. 
          The world pion data are taken from Refs.~\cite{pion_exp1,pion_exp2,pion_exp3,pion_exp4,pion_exp5,pion_exp6,pion_exp7,pion_exp8}.}
      \label{fig:pi_ff_total}
\end{figure}

\begin{figure}[t]
    \hspace{-1.8cm}\begin{tabular}{cc}
      \resizebox{8.9cm}{!}{\includegraphics{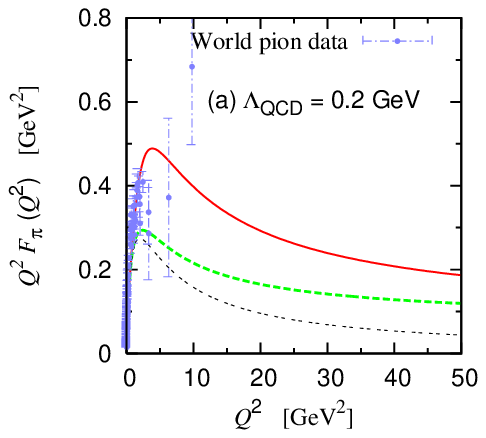}} &
      \resizebox{8.9cm}{!}{\includegraphics{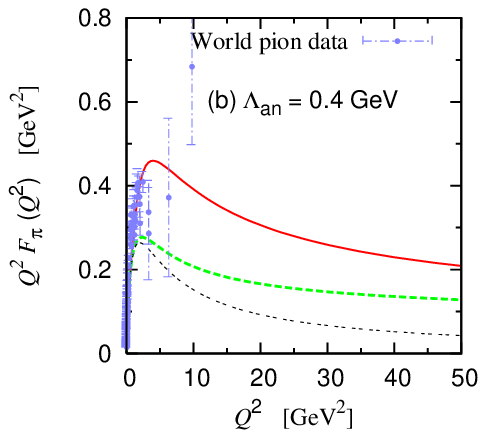}} \\
    \end{tabular}
    \caption{Scaled total electromagnetic form factor for the pion with (a) the
         usual QCD running coupling and (b) the analytical running
         coupling over a wider range of the intermediate energies. The solid 
         line represents the full twist-3 result (Soft+Twist-2 +Twist-3), the 
         long dashed lines represent the full twist-2 result (Soft+Twist-2) 
         and the soft corrections are indicated by the short dashed lines. The 
         world pion data are taken from Refs.~\cite{pion_exp1,pion_exp2,pion_exp3,pion_exp4,pion_exp5,pion_exp6,pion_exp7,pion_exp8}.}
    \label{fig:pi_ff_total_long}
\end{figure}

Our final results for the scaled pion form factor are summarized in
Fig.~\ref{fig:pi_ff_total} and Fig.~\ref{fig:pi_ff_total_long}. We use both 
the usual and the analytic QCD running couplings and compare our results with 
the available experimental pion data with increasing error bars towards
intermediate energies. The individual soft and hard contributions along with 
the total contribution are shown. Clearly, contrary to the earlier claims 
made in Ref.~\cite{pff3}, the twist-2 hard form factor is far too small in the 
phenomenologically accessible region to explain the data. One must therefore 
look for other possibilities like non-perturbative higher twist effects and  
soft contributions. Interestingly, it is seen that the soft dynamics largely 
dominate the low-energy region below 10-16 GeV$^2$ but rapidly fall off in 
the asymptotic region. The contributions from the twist-3 corrections turn out 
to be significantly large in the moderate range of energies below 100
GeV$^2$, but eventually the hard twist-2 contributions solely determine the 
asymptotic trend beyond $Q^2\approx$ 100-150 GeV$^2$.    

\begin{figure}[t]
\begin{center}  
      \resizebox{14.1cm}{!}{\includegraphics{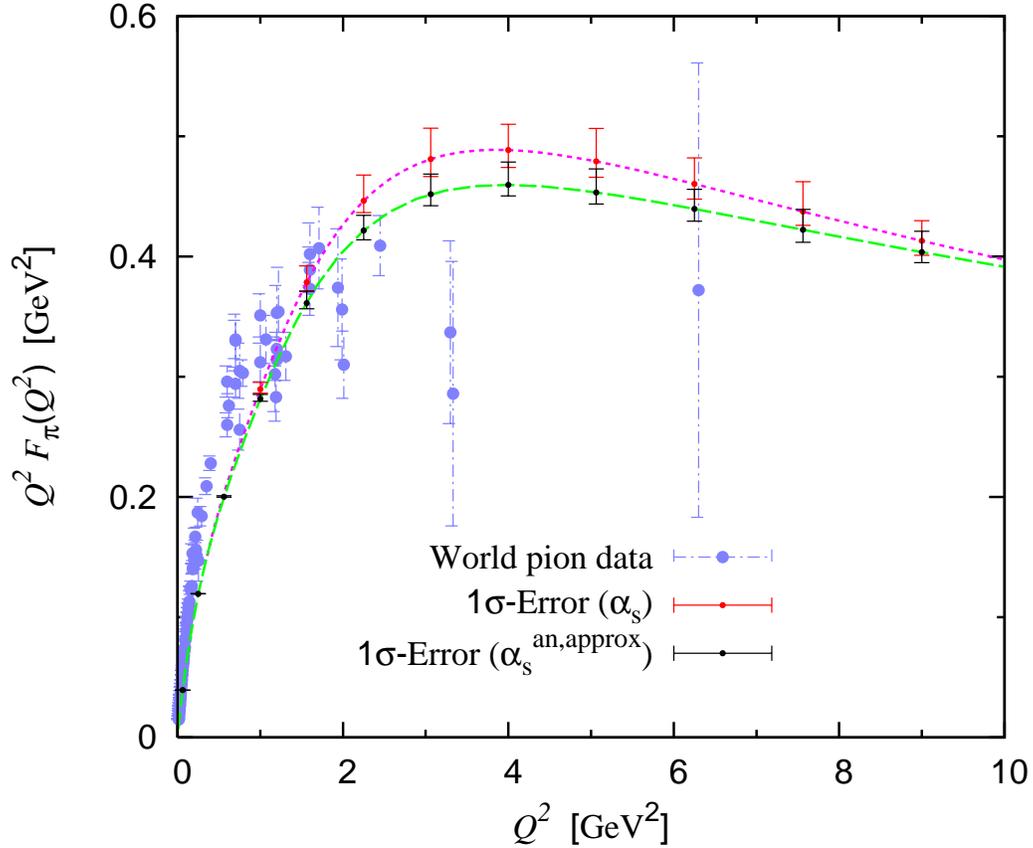}}
      \caption{Theoretical $1\sigma$-error for the scaled total pion form 
        factor due to the variation of the input parameters for the DAs 
        ($\chi_{3\pi},\,f_{3\pi},\,\omega_{3\pi}$ and $a^{\pi}_2$ ) and
        $\Lambda_{\rm QCD,an}$. The lines correspond to the mean 
        values of the parameters using the usual (dotted line) and the
        analytic (dashed line) QCD coupling schemes, respectively. The
        error bars for the experimental data points
        \cite{pion_exp1,pion_exp2,pion_exp3,pion_exp4,pion_exp5,pion_exp6,pion_exp7,pion_exp8} are also shown.}
      \label{fig:pi_ff_total_gauss}
\end{center}  
\end{figure}

As evident from the figures, the total scaled pion form factor (solid lines) up
to twist-3 corrections displays an obvious improvement of the overall 
agreement with experimental data compared to the twist-2 scaled form factor 
(dashed lines). To some extent, it is somewhat surprising to see that in 
combination with the soft contributions, the modified resummed pQCD with the 
usual QCD coupling could work so well as low as  $Q^2\approx 0.25$ GeV$^2$, far
lower than previously envisaged. To this end, we display the results in the 
analytic scheme to confirm our results. The analytical prescription is known to
reduce the scheme and renormalization scale dependence, largely increasing the 
stability of solutions \cite{pff6}. Accordingly, there is some confidence in 
our displayed results. The results obtained in both schemes not only show a 
striking similarity even at sufficiently low momentum transfers, but also show 
a good agreement with the available pion form factor data. However, whether 
or not such an agreement is merely accidental is matter of debate. 
There may still be substantial sub-leading contributions e.g., from a full 
NLO calculation in the strong coupling including subleading twists and 
intrinsic transverse momenta for the hard scattering kernel and the DAs, or 
from higher order Fock states and helicity components in the light-cone DAs. 
Note that a NLO calculation in the strong QCD coupling constant was done in 
Ref.~\cite{pff5} where the corrections to the twist-2 pion form factor 
was found to be quite large. A full NLO calculation for subleading twists, 
including also the transverse momentum dependence, is however, still missing. 
In Ref.~\cite{pff8}, contributions of higher helicity states were 
found to lower the total pion form factor significantly. Hence, without 
systematically taking all of these effects in account, which is beyond the 
scope of the paper, no definitive statement can be made as to how well our 
results agree with the data. Moreover, the available data itself has very low 
statistics at intermediate energy and is plagued with large uncertainties. It 
is therefore difficult to give a proper theoretical error estimate of our 
results, when the asymptotic formalism is itself largely unreliable in the 
region of our interest. What we have done in this paper is a combination of 
model and pQCD calculation. It may thus be worth using the estimates for the 
ranges over which the input parameters of the DAs, namely 
$\mu_\pi\,(\chi_{3\pi}),\,f_{3\pi},\,\omega_{3\pi}$ and $a^{\pi}_2$ vary (given
in the Tables~\ref{table:para} \& \ref{table:para-det}), in computing our 
theoretical error. In addition, we allow a variation of $\pm 0.05$ GeV for both
$\Lambda_{\rm QCD}$ and $\Lambda_{\rm an}$. We used a Monte Carlo technique to 
generate a gaussian ``$1\sigma$'' spread of the scaled form factor for various 
$Q^2$ values about a central mean. The error estimate displayed in 
Fig.~\ref{fig:pi_ff_total_gauss} shows our maximum theoretical error to be 
about 10 percent for the usual QCD coupling and somewhat less for the analytic 
coupling.

\begin{figure}
\begin{center}
      \resizebox{13.5cm}{!}{\includegraphics{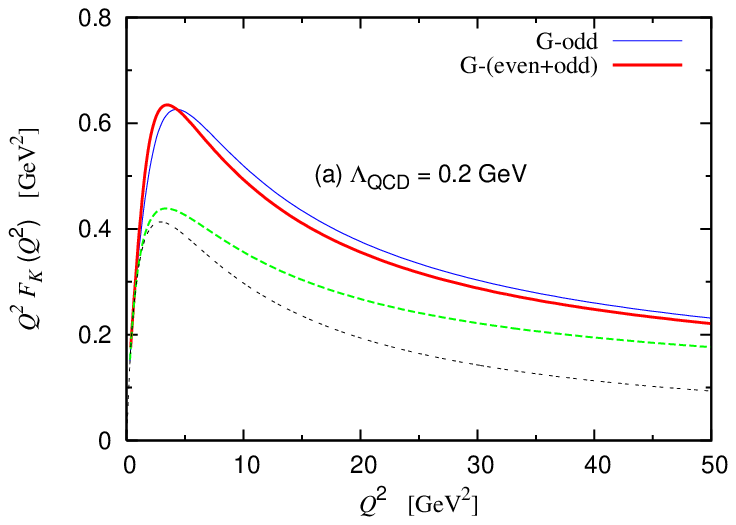}}\\
      \resizebox{13.5cm}{!}{\includegraphics{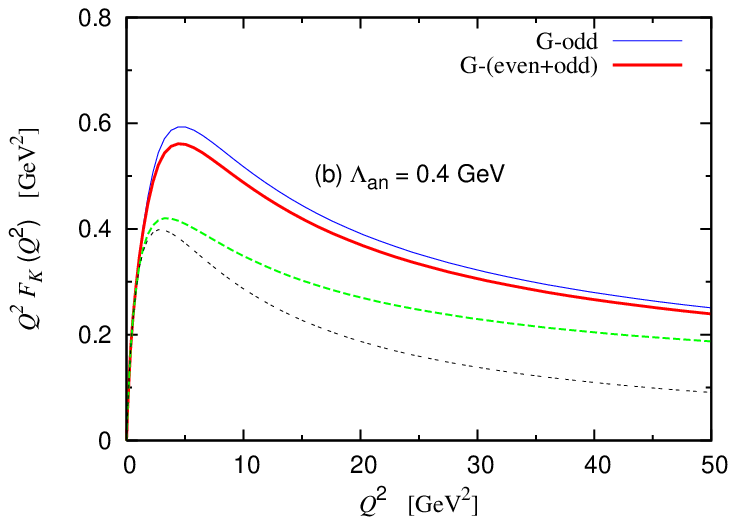}}
      \caption{Scaled total electromagnetic form factor for the kaon with the
      usual QCD running coupling (upper plot) and the analytical running
      coupling (lower plot) for intermediate energies. The solid lines 
      represents the full twist-3 result (Soft+Twist-2 +Twist-3), the long 
      dashed lines represent the twist-2 result (Soft+Twist-2) and the 
      soft corrections are indicated by the short dashed lines. For the full
      twist-3 case, both the results i.e., with and without including the
      G-parity-breaking terms in the DAs are shown.}
      \label{fig:K_ff_total}
\end{center}  
\end{figure} 
               
\subsection{The kaon form factor}
\label{sec:kaon}
We conclude the section on the numerical analysis by displaying our predictions
for the kaon form factor, applying the same techniques as for the case of the 
pion form factor. The twist-2 and twist-3 light-cone wave functions for the 
kaon has the general form in the transverse $b$-space given by
\begin{eqnarray}
\tilde{{\mathcal P}}_K(x,b,1/b,{\mathcal
  M}_{u,d,s})\!&=&\!A_K\,\phi_K(x,1/b)\,{\rm{exp}}\left[-\frac{b^2\,x(1-x)}{4\beta^2_K}\right]\nonumber\\
&&\times{\rm{exp}}\left[-\beta^2_K\left(\frac{{\mathcal M}^2_{s}}{x}+\frac{{\mathcal M}^2_{u,d}}{1-x}\right)\right]\,,
\end{eqnarray}
where we used the twist-3 chiral enhancement parameter $\chi_{3K}=1.3$ GeV
and the experimental estimate for the kaon decay constant $f_K\approx 1.22
f_\pi$ \cite{wfs0}. Here, we also take $\Lambda_{\rm QCD}=0.2$ GeV and 
$\Lambda_{\rm an}=0.4$ GeV for the respective running couplings in the 
$\overline{MS}$ scheme. The results are summarized in Fig.~\ref{fig:K_ff_total}
for intermediate energies. The solid line represents the total scaled form 
factor in each case. Here, our results for the kaon form factor must be 
considered preliminary. Due to the complete absence of experimental data at 
intermediate energies we are unable to make any meaningful phenomenological 
comparison. The presently available kaon data has very poor statistics and has 
hardly been measured above 0.2 GeV$^2$. Hence, we do not show the experimental 
data points in the form factor plots. With the availability of better quality 
data in future there could be plenty of room for further improvements, for 
instance, extension of the above results to include a full NLO calculation for 
subleading twists or higher helicity and Fock state contributions.

\section{Discussion and Conclusion}
\label{sec:concl}
For the past two decades the electromagnetic meson form factors have been
the subject of intensive theoretical and experimental scrutiny and yet there
is still to be an universally accepted framework for their description.
Presently, reliable experimental data are available only for the pions which 
are entirely concentrated at very low energies with very poor statistics at 
intermediate energies. The low energy part of the data is best explained by the
standard VMD model, showing no apparent trace of pQCD behavior, which is 
expected only at very high energies. Very many attempts have been made to 
predict the onset of the perturbative behavior for the pion form factor. The 
modified or resummed valence pQCD with factorization appears to show some 
attractive features to enable pQCD calculations to be valid in a 
self-consistent way even at very moderate energies. Whether this is true, will 
only be confirmed when data with better statistics at higher momentum transfers
will become available in the future. At the same time, the onset of the 
perturbative behavior being very slow, it is still unclear whether the leading 
order perturbative calculations could be expected to be precise even at the 
highest accessible energies. In this paper, with the help of (a) the 
double-humped type DAs and (b) the modified transverse ($k_T$) factorization 
scheme, incorporating both Sudakov suppression and threshold resummation, we 
got rid of non-perturbative end-point enhancements. This enlarges the scope of 
applicability of resummed pQCD independent of the coupling scheme to a much 
wider range of intermediate energies, if not down to a few GeVs, as 
demonstrated in this paper. By a simple adjustment of only the chiral 
enhancement parameter $\chi_{3\pi}$, a good agreement with the experimental 
data was obtained.

As for the scaled pion form factor, we found that the leading order 
pQCD contributions potentially undermine the agreement with the available low 
energy data and are only trustworthy in the hard energy regime: even from a 
very conservative point of view, $Q^2$ should be bigger than 4 GeV$^2$. At low 
momentum transfers, the non-perturbative contributions dominate, being larger 
than the hard (twist-2) contributions at least by a factor of two. In fact 
below 4 GeV$^2$, 60-70$\%$ of the available data is already accounted for by 
the soft contributions. In addition, we also needed the twist-3 power 
corrections to explain the remaining discrepancy. However, at larger energies 
(say, $Q^2>50$-100 GeV$^2$), both the soft and the twist-3 contributions 
rapidly fall off and eventually the twist-2 form factor dominates 
asymptotically. Similar conclusions, albeit being preliminary, are drawn for 
the kaon form factor although it seems that the onset of the perturbative 
behavior occurs at slightly larger momentum transfers than for the case of the 
pion. Of course, as we mentioned earlier, it still remains to be investigated 
about the nature of the contributions that may arise from a full systematic NLO
calculation with subleading twists and intrinsic transverse momenta, or from 
the inclusion of higher helicity and Fock states. With the availability of 
better quality of data in future such analyses may be necessary to make 
definite conclusions.

In summary, although the quality of present experimental form factor 
data does not allow a definitive conclusion, one can expect that the 
non-perturbative soft contributions and higher twist power corrections to the 
form factors play an important role at phenomenologically accessible 
momentum transfers. Thus, more work is needed to be done on both the 
theoretical and experimental sides to obtain more conclusive results and push 
the frontiers of our knowledge on confinement dynamics through the study of 
higher order and non-perturbative contribution to exclusive processes.\\

\noindent {\it Acknowledgments:} The authors would like to thank Alexander 
Bakulev, Pankaj Jain, Alexander Lenz, Seregy Mikhailov, Sabyasachi Mishra, Ingo
Sick, Ritesh Singh, Nikolaos Stefanis and Xing-Gang Wu for interesting 
discussions and comments. One of the authors, U.R. would like to convey special
thanks to Prof. Dirk Trautmann and Prof. Friedel Thielemann for their 
hospitality and financial support at the University of Basel. This work was 
supported by the Swiss National Science Foundation (SNSF) under the contract 
number 200020-111705. \\ 

\pagebreak


\appendix
\renewcommand{\thesection}{\Alph{section}}
\renewcommand{\theequation}{\Alph{section}\arabic{equation}}

\setcounter{equation}{0}

\section{2-particle twist-3 Distribution Amplitudes}
\label{app:DA}
The twist-3 DAs are obtained by an expansion over conformal spins. At
next-to-leading order, the 2-particle DAs $\phi^p_{3,M}$ and 
$\phi^\sigma_{3,M}$ (including meson-mass corrections that break chiral 
symmetry at ${\mathcal O}(m_s+m_q)$ in the SU(3) case while preserving 
G-parity) are given in terms the Gegenbauer polynomials $C^{1/2}_n$ and  
$C^{3/2}_n$, respectively \cite{DAs1,DAs2} as
\begin{eqnarray}
\phi^p_{3;M}(x,\mu^2)\!&=&\!\frac{f_M}{4\sqrt{2N_c}}\left\{1+\left(30\eta_{3M}(\mu^2)-\frac{5}{2}\,\rho^2_M(\mu^2)\right)C^{1/2}_2(\xi)\right.\nonumber\\
&&+\left.\left(-3\eta_{3M}(\mu^2)\,\omega_{3M}(\mu^2)-\frac{27}{20}\,\rho^2_M(\mu^2)-\frac{81}{10}\,\rho^2_M(\mu^2)\,a^{M}_2(\mu^2)\right)C^{1/2}_4(\xi)\right\}\,,\nonumber
\end{eqnarray}
\begin{eqnarray}
\phi^\sigma_{3;M}(x,\mu^2)\!&=&\!\frac{3f_M}{2\sqrt{2N_c}}x(1-x)\left\{1+\left(5\eta_{3M}(\mu^2)-\frac{1}{2}\eta_{3M}(\mu^2)\,\omega_{3M}(\mu^2)\right.\right.\nonumber \\
&&-\left.\left.\frac{7}{20}\,\rho^2_M(\mu^2)-\frac{3}{5}\,\rho^2_M(\mu^2)
   \,a^{M}_2(\mu^2)\right)C^{3/2}_2(\xi)\right\}
\end{eqnarray}
with
\begin{equation}
\eta_{3M}=\frac{f_{3M}}{f_M}\frac{1}{\mu_M}\quad ;\quad
\rho_M=\frac{m_M}{\mu_M}\,\,;\quad M=\pi^{\pm},K^{\pm}
\end{equation} 
where the non-perturbative parameter $f_{3M}$ and $\omega_{3M}$, 
respectively are defined by the following matrix elements of local twist-3 
operators:
\begin{eqnarray}
\left<0\left|{\bar q}_{f_{1}}\sigma_{\mu\nu}\gamma_5\, g_sG_{\alpha\beta}q_{f_2}
\right|M(P)\right>=if_{3M}\left( P_\alpha P_\mu g_{\nu\beta}-
P_\alpha P_\nu g_{\mu\beta} - P_\beta P_\mu g_{\nu\alpha}+ P_\beta P_\nu
g_{\alpha\mu}\right) \,, \nonumber
\end{eqnarray}
\begin{eqnarray}
\left<0|{\bar
    q}_{f_1}\sigma_{\mu\lambda}\gamma_5[iD_\beta,g_sG_{\alpha\lambda}]q_{f_2}\right.\!\!&-&\!\!\left.(3/7)\,i\partial_\beta\,{\bar q}_{f_1}\sigma_{\mu\lambda}\gamma_5\,g_sG_{\alpha\lambda}q_{f_2}|M(P)\right>\nonumber \\
&=&\!\frac{3}{14}\,if_{3M}P_\alpha P_\beta P_\mu\,\omega_{3M}\,,
\end{eqnarray}
where $G_{\alpha\beta}$ is the gluon field tensor. The LO scale dependence of 
various twist-3 parameters are given by
\begin{eqnarray}
\rho_{M}(\mu^2)\!&=&\!L^{\gamma^{(0)}_{3;q{\bar q}}/\beta_0}\,\rho_{M}(\mu^2_0)\,;
\quad\gamma^{(0)}_{3;q{\bar q}}=1\,, \nonumber\\
\mu_{M}(\mu^2)\!&=&\!L^{\gamma^{(0)}_{3;\mu}/\beta_0}\,\mu_{M}(\mu^2_0)\,;\quad \gamma^{(0)}_{3;\mu}=-\gamma^{(0)}_{3;q{\bar q}}=-1\,,\nonumber \\
\eta_{3M}(\mu^2)\!&=&\!L^{\gamma^{(0)}_{3;\eta}/\beta_0}\,\eta_{3M}(\mu^2_0)\,;\quad
\gamma^{(0)}_{3;\eta}=\frac{4}{3}\,{\mathcal C}_F + \frac{1}{4}\,{\mathcal C}_A
\,,\nonumber \\
\omega_{3M}(\mu^2)\!&=&\!L^{\gamma^{(0)}_{3;\omega}/\beta_0}\,\omega_{3M}(\mu^2_0)\,;\quad
\gamma^{(0)}_{3;\omega}=-\frac{7}{24}\,{\mathcal C}_F + \frac{7}{12}\,{\mathcal C}_A\,,\nonumber\\
a^{M}_2(\mu^2)\!&=&\!L^{\gamma^{(0)}_2/\beta_0}\,a^{M}_2(\mu^2_0)\,;\quad
\gamma^{(0)}_2= \frac{25}{24}\,{\mathcal C}_F\,,
\end{eqnarray}
where $L=\alpha_s(\mu^2)/\alpha_s(\mu^2_0)$,\, ${\mathcal
  C}_F=(N^2_c-1)/2N_c$, ${\mathcal C}_A=N_c$ and $\mu_0\approx1$ GeV.

We have also considered the 2-particle twist-3 kaon DAs 
$\phi^{\,p}_{3,K}$ and $\phi^\sigma_{3,K}$, as given in \cite{DAs4}, which not
only include a complete set of meson-mass corrections but also
G-parity-breaking terms of ${\mathcal O}(m_s-m_q)$:
\begin{eqnarray}
\phi^p_{3;K}\!\!\!&\,&\!\!\!\!\!\!\!(x,\mu^2)=\frac{f_K}{4\sqrt{2N_c}}\left\{1+3\rho^{K}_+(1+6a^K_2)-9\rho^{K}_-a^K_1+\left[\,\frac{27}{2}\,\rho^K_+a^K_1\right.\right.\nonumber\\
&-&\!\!\!\left.\rho^K_-\left(\frac{3}{2}+27a^K_2\right)\right]C^{1/2}_1(\xi)+\left(30\eta_{3K}+15\rho^K_+a^K_2-3\rho^K_-a^K_1\left)C^{1/2}_2(\xi) \right.\right.\nonumber \\
&+&\!\!\!\left(10\eta_{3K}\lambda_{3K}-\frac{9}{2}\rho^K_-a^K_2\right)C^{1/2}_3(\xi)
-3\eta_{3K}\omega_{3K}C^{1/2}_4(\xi)+\frac{3}{2}\,\left(\rho^K_+
+\rho^K_-\right)\nonumber\\
&\times&\!\!\!\left.\left(1-3a^K_1+6a^K_2\right)\ln x
  +\frac{3}{2}\,\left(\rho^K_+-\rho^K_-\right)\left(1+3a^K_1+6a^K_2\right)\ln
  (1-x)\right\}\,, \nonumber\\
&\,&
\end{eqnarray}

\vspace{-1cm}

\begin{eqnarray}
\phi^\sigma_{3;K}\!\!\!&\,&\!\!\!\!\!\!\!(x,\mu^2)=\frac{3f_K}{2\sqrt{2N_c}}\,x(1-x)\left\{1+\frac{3}{2}\,\rho^K_++15\rho^K_+a^K_2-\frac{15}{2}\,\rho^K_-a^K_1\right.\nonumber\\
&+&\!\!\!\left(3\rho^K_+a^K_1-\frac{15}{2}\,\rho^K_-a^K_2\right)C^{3/2}_1(\xi)+\left(5\eta_{3K}-\frac{1}{2}\,\eta_{3K}\omega_{3K}+\frac{3}{2}\,\rho^K_+a^K_2\right)C^{3/2}_2(\xi)\nonumber\\
&+&\!\eta_{3K}\lambda_{3K}C^{3/2}_3(\xi)+\frac{3}{2}\,\left(\rho^K_++\rho^K_-\right)\left(1-3a^K_1+6a^K_2\right)\ln x \nonumber\\ 
&+&\!\!\!\left.\frac{3}{2}\,\left(\rho^K_+-\rho^K_-\right)\left(1+3a^K_1+6a^K_2\right)\ln (1-x)\right\}\left(\frac{1}{1-\rho_+}\right)
\end{eqnarray}
with
\begin{equation}
\eta_{3K}=\frac{f_{3K}}{f_K}\frac{1}{\mu_K}\,\,,\,\,
\rho^K_+=\frac{(m_s+m_q)^2}{m^2_K}\,\,\,\,{\rm
  and}\,\,\,\,\rho^K_-=\frac{m^2_s-m^2_q}{m^2_K}\,. 
\end{equation} 
Note that the expression for $\phi^\sigma_{3;K}(x,\mu^2)$ is normalized to 
unity with an extra factor of $1/(1-\rho_+)$, compared to that given in 
\cite{DAs4}. The non-perturbative parameters $f_{3K}$, $\omega_{3K}$ and  
$\lambda_{3K}$ are defined (e.g., $K^-$) by
\begin{eqnarray}
\left<0\left|{\bar u}\,\sigma_{\mu\nu}\gamma_5\, g_sG_{\alpha \beta}\,
    s\,\right|K^-(P)\right> = if_{3K}\left( P_\alpha P_\mu
    g_{\nu\beta}-P_\alpha P_\nu
    g_{\mu\beta}\right.-\left.P_\beta P_\mu
    g_{\nu\alpha}+P_\beta P_\nu g_{\alpha\mu}\right) \,, \nonumber
\end{eqnarray}

\vspace{-0.6cm}

\begin{eqnarray}
\left<0\left|{\bar u}\,\sigma_{\mu\lambda}\gamma_5[iD_\beta,g_sG_{\alpha
\lambda}]s-(3/7)\,i\partial_\beta\,{\bar u}\,\sigma_{\mu\lambda}
\gamma_5\, g_sG_{\alpha \lambda}s\,\right|K^-(P)\right> =
\frac{3}{14}\,if_{3K}P_\alpha P_\beta P_\mu\,\omega_{3K}\,,\nonumber
\end{eqnarray}

\vspace{-0.6cm}

\begin{eqnarray}
\label{eq:fwl_3K}
\left<0\left|{\bar u}\,i\overleftarrow{D}_{\beta}\sigma_{\mu\lambda}\gamma_5 
g_sG_{\alpha\lambda}\,s-{\bar u}\,\sigma_{\mu\lambda}\gamma_5 g_sG_{\alpha
\lambda}i\overrightarrow{D}_{\beta}\,s\,\right|K^-(P)\right>\!&=&\!
\frac{1}{7}\,if_{3K}P_\alpha P_\beta P_\mu\lambda_{3K}\,.\nonumber\\
& &
\end{eqnarray}
where in the chiral limit the renormalization group equations at LO give,
\begin{eqnarray}
\mu_{K}(\mu^2)\!&=&\!L^{\gamma^{(0)}_{3;\mu}/\beta_0}\,\mu_{K}(\mu^2_0)\,;
\quad \gamma^{(0)}_{3;\mu}=-\gamma^{(0)}_{3;q{\bar q}}=-1\,,\nonumber \\
\rho^K_+(\mu^2)\!&=&\!L^{\gamma^{(0)}_{3;\rho^+}/\beta_0}\,\rho^K_+(\mu^2_0)\,;\quad \gamma^{(0)}_{3;\rho^+}=2\gamma^{(0)}_{3;q{\bar q}}=2\,,\nonumber\\
\rho^K_-(\mu^2)\!&=&\!L^{\gamma^{(0)}_{3;\rho^-}/\beta_0}\,\rho^K_-(\mu^2_0)\,;\quad \gamma^{(0)}_{3;\rho^-}=2\gamma^{(0)}_{3;q{\bar q}}=2\,,\nonumber\\
f_{3K}(\mu^2)\!&=&\!L^{\gamma^{(0)}_{3;f}/\beta_0}\,f_{3K}(\mu^2_0)\,;\quad
\gamma^{(0)}_{3;f}=\frac{7}{12}\,{\mathcal C}_F + \frac{1}{4}\,{\mathcal C}_A
\,,\nonumber \\
\omega_{3K}(\mu^2)\!&=&\!L^{\gamma^{(0)}_{3;\omega}/\beta_0}\,\omega_{3K}(\mu^2_0)\,;\quad
\gamma^{(0)}_{3;\omega}=-\frac{7}{24}\,{\mathcal C}_F + \frac{7}{12}\,{\mathcal
  C}_A\,,\nonumber \\
\lambda_{3K}(\mu^2)\!&=&\!L^{\gamma^{(0)}_{3;\lambda}/\beta_0}\,\lambda_{3K}(\mu^2_0)\,;\quad
\gamma^{(0)}_{3;\lambda}=\frac{19}{48}\,{\mathcal C}_F\,, \nonumber\\
a^{K}_1(\mu^2)\!&=&\!L^{\gamma^{(0)}_1/\beta_0}\,a^{K}_1(\mu^2_0)\,;
\quad \gamma^{(0)}_1= \frac{2}{3}\,{\mathcal C}_F\,, \nonumber \\
a^{K}_2(\mu^2)\!&=&\!L^{\gamma^{(0)}_2/\beta_0}\,a^{K}_2(\mu^2_0)\,;
\quad \gamma^{(0)}_2= \frac{25}{24}\,{\mathcal C}_F\,,
\end{eqnarray}
But the strange quark being massive, there is operator mixing of the ones in
Eq.~(\ref{eq:fwl_3K}) with that of twist-2 operators and the resulting LO 
renormalization group equations give the following scale dependence of the 
various twist-3 parameters:
\begin{eqnarray}
f_{3K}(\mu^2)\!&=&\!L^{55/(36\beta_0)}f_{3K}(\mu^2_0)+\frac{2}{19}\left(L^{1/\beta_0}-L^{55/(36\beta_0)}\right)[f_Km_s](\mu^2_0)\nonumber\\
&&+\,\frac{6}{65}\left(L^{55/(36\beta_0)}-L^{17/(9\beta_0)}\right)[f_Km_sa^K_1](\mu^2_0)\,,\nonumber\\
\left[f_{3K}\,\omega_{3K}\right](\mu^2)\!&=&\!L^{26/(9\beta_0)}[f_{3K}\,\omega_{3K}](\mu^2_0)+\frac{1}{170}\left(L^{1/\beta_0}-L^{26/(9\beta_0)}\right)[f_Km_s](\mu^2_0)\nonumber\\
&&+\,\frac{1}{10}\left(L^{17/(9\beta_0)}-L^{26/(9\beta_0)}\right)[f_Km_sa^K_1](\mu^2_0)\nonumber\\
&&+\,\frac{2}{15}\left(L^{43/(18\beta_0)}-L^{26/(9\beta_0)}\right)[f_Km_sa^K_2](\mu^2_0)\,,\nonumber\\
\left[f_{3K}\lambda_{3K}\right](\mu^2)\!&=&\!L^{37/(18\beta_0)}[f_{3K}\lambda_{3K}](\mu^2_0)-\frac{14}{67}\left(L^{1/\beta_0}-L^{37/(18\beta_0)}\right)[f_Km_s](\mu^2_0)\nonumber\\
&&+\,\frac{14}{5}\left(L^{17/(9\beta_0)}-L^{37/(18\beta_0)}\right)[f_Km_sa^K_1](\mu^2_0)\nonumber\\
&&-\,\frac{4}{11}\left(L^{43/(18\beta_0)}-L^{37/(18\beta_0)}\right)[f_Km_sa^K_2](\mu^2_0)\,.
\end{eqnarray}
Finally, we present the various Gegenbauer polynomials used in the above 
formulae:
\begin{eqnarray}
C^{1/2}_1(\xi)\!&=&\!\xi\,,\nonumber\\
C^{1/2}_2(\xi)\!&=&\!\frac{1}{2}\,(3\xi^2-1)\,,\nonumber\\
C^{1/2}_3(\xi)\!&=&\!\frac{1}{2}\,\xi(5\xi^2-3)\,,\nonumber\\
C^{1/2}_4(\xi)\!&=&\!\frac{1}{8}\,(35\xi^4-30\xi^2+3)\,,\nonumber\\
C^{3/2}_0(\xi)\!&=&\!1\,,\nonumber\\
C^{3/2}_1(\xi)\!&=&\!3\xi\,,\nonumber\\
C^{3/2}_2(\xi)\!&=&\!\frac{3}{2}\,(5\xi^2-1)\,,\nonumber\\
C^{3/2}_3(\xi)\!&=&\!\frac{5}{2}\,\xi(7\xi^2-3)\,.
\end{eqnarray}
\section{Calculation of the Sudakov exponent}
\label{app:sudakov}
The full expression of the Sudakov suppression factor $S(x,y,b_1,b_2,Q)$ is 
given by,
\begin{eqnarray}
S(x,y,b_1,b_2,Q)\!&=&\!s(xQ,b_1)+s(yQ,b_2)+s((1-x)Q,b_1)+s((1-y)Q,b_2) \nonumber\\
&&-\,\frac{1}{\beta_0}\ln\left(\frac{\hat{t}}{-\hat{b}_1}\right)-\frac{1}
{\beta_0}\ln\left(\frac{\hat{t}}{-\hat{b}_2}\right)\nonumber\\
&&+\,\frac{\beta_1}{\beta^3_0}\left[\frac{1+\ln(-2\hat{b}_1)}{-2\hat{b}_1}-\frac{1+\ln(2\hat{t})}{2\hat{t}}\right]\nonumber\\
&&+\,\frac{\beta_1}{\beta^3_0}\left[\frac{1+\ln(-2\hat{b}_2)}{-2
\hat{b}_2}-\frac{1+\ln(2\hat{t})}{2\hat{t}}\right]\,,
\end{eqnarray}
where
\begin{eqnarray}
s(XQ,1/b)\!&=&\!\frac{{\mathcal A}^{(1)}}{2\beta_0}\,\hat{q}\ln\left(\frac{\hat{q}}{-\hat{b}}\right)+\frac{{\mathcal A}^{(2)}}{4\beta^2_0}\left(\frac{\hat{q}}{-\hat{b}}-1\right)-\frac{{\mathcal A}^{(1)}}{2\beta_0}\left(\hat{b}+\hat{q}\right)\nonumber\\
&&-\,\left[\frac{4{\mathcal A}^{(1)}\beta_1}{16\beta^3_0}\,\hat{q}+\frac{2{\mathcal A}^{(1)}\beta_1}{16\beta^3_0}\ln\left(\frac{1}{2}e^{2\gamma_{E}-1}\right)\right]\left[\frac{1+\ln(-2\hat{b})}{-\hat{b}}-\frac{1+\ln(2\hat{q})}{\hat{q}}\right]\nonumber\\
&&-\,\left[\frac{{\mathcal A}^{(2)}}{4\beta^2_0}-\frac{{\mathcal A}^{(1)}}{4\beta_0}\ln\left(\frac{1}{2}e^{2\gamma_{E}-1}\right)\right]\ln\left(\frac{\hat{q}}{-\hat{b}}\right)-\frac{4{\mathcal A}^{(1)}\beta_1}{32\beta^3_0}\left[\ln^2(-2\hat{b})-\ln^2(2\hat{q})\right]\nonumber\\
&&+\,\frac{2{\mathcal A}^{(2)}\beta_1}{8\beta^4_0}\left[\frac{1+\ln(-2\hat{b})}{-\hat{b}}-\frac{1+\ln(2\hat{q})}{\hat{q}}\right] \nonumber\\
&&-\,\frac{2{\mathcal A}^{(2)}\beta_1}{8\beta^4_0}\,\hat{q}\left[\frac{1+2\ln(-2\hat{b})}{(-2\hat{b})^2}-\frac{1+2\ln(2\hat{q})}{(2\hat{q})^2}\right]\nonumber\\
&&-\,\frac{4{\mathcal A}^{(2)}\beta^2_1}{64\beta^6_0}\left[\frac{1+2\ln(-2\hat{b})+2\ln^2(-2\hat{b})}{(-2\hat{b})^2}-\frac{1+2\ln(2\hat{q})+2\ln^2(2\hat{q})}{(2\hat{q})^2}\right]\nonumber\\
&&+\,\frac{4{\mathcal A}^{(2)}\beta^2_1}{8\beta^6_0}\,\hat{q}\left[\frac{\frac{2}{27}+\frac{2}{9}\ln(-2\hat{b})+\frac{1}{3}\ln^2(-2\hat{b})}{(-2\hat{b})^3}-\frac{\frac{2}{27}+\frac{2}{9}\ln(2\hat{q})+\frac{1}{3}\ln^2(2\hat{q})}{(2\hat{q})^3}\right]\,. \nonumber\\
& &
\end{eqnarray}
In the above formulae,
\begin{eqnarray}
\hat{t}\!&=&\!\ln\left(\frac{t}{\Lambda_{\rm QCD}}\right);\,\,t={\rm max}(\sqrt{xy}\,Q,1/b_1,1/b_2)\,,\nonumber\\
\hat{b}\!&=&\!\ln\left(b\,\Lambda_{\rm QCD}\right)\,,\nonumber\\
\hat{q}\!&=&\!\ln\left[\frac{XQ}{\sqrt{2}\Lambda_{\rm
      QCD}}\right];\,\,X=x,y,(1-x)\,\,{\rm or},\,(1-y)\,,
\nonumber\\
{\mathcal A}^{(1)}\!&=&\!{\mathcal C}_F=\frac{4}{3}\nonumber\,,\\
{\mathcal A}^{(2)}\!&=&\!\left(\frac{67}{27}-\frac{\pi^2}{9}\right)N_c-\frac{10}{27}N_f+\frac{8}{3}\beta_0\ln\left(\frac{e^{\gamma_{E}}}{2}\right)\,.
\end{eqnarray}




\begin{thebibliography}{}

\bibitem{pion_exp1}
C.~N.~Brown et al.,
Phys.\ Rev.\ D {\bf 8} (1973) 92.

\bibitem{pion_exp2}
C.~J.~Babek et al.,
Phys.\ Rev.\ D {\bf 17} (1978) 1693.

\bibitem{pion_exp3}
H.~Ackermann et al.,
Nucl.\ Phys.\ B {\bf 137} (1978) 294.

\bibitem{pion_exp4}
P.~Brauel et al.,
Z.\ Phys.\ C {\bf 3} (1979) 101.

\bibitem{pion_exp5}
S.~R.~Amendolia et al.,
Phys.\ Lett.\ B {\bf 178} (1986) 435; 
Phys.\ Lett.\ B {\bf 277} (1986) 168. 

\bibitem{pion_exp6}
J.~Volmer,
Ph.D. thesis, Vrije Universiteit, Amsterdam, 2000 (unpublished);
Phys.\ Rev.\ Lett. {\bf 86} (2001) 1713.

\bibitem{pion_exp7}
T.~Horn et al.,
Phys.\ Rev.\ Lett. {\bf 97} (2006) 192001.

\bibitem{pion_exp8}
V.~Tadevosyan et al.,
Phys.\ Rev.\ C {\bf 75} (2007) 055205.

\bibitem{kaon_exp1}
B.~Zeidman et al.,
CEBAF Experiment E91-016/1996.

\bibitem{kaon_exp2}
R.~Mohring et al., 
Phys.\ Rev.\ Lett. {\bf 81} (1998) 1805.

\bibitem{kaon_theo1}
M.~Vanderhaeghen, M.~Guidal and J.-M.~Laget, 
Phys.\ Rev.\ C {\bf 57} (1998) 1454.

\bibitem{kaon_theo2}
C.~Bennhold and T.~Mart, 
Phys.\ Rev.\ C {\bf 61} (1999) 012201.

\bibitem{Williams}
R.~A.~Williams, 
Phys.\ Rev.\ C {\bf 46} (1992) 1617.

\bibitem{Dally}
E.~B~.~Dally et al.,
Phys.\ Rev.\ Lett. {\bf 45} (1980) 232.

\bibitem{Glander}
K.~H.~Glander et al., 
Eur.\ Phys.\ J.\ A {\bf 19} (2004) 251.
 
\bibitem{McNabb}
J.~W.~C.~McNabb et al., 
Phys.\ Rev.\ C {\bf 69} (2004) 042201(R).

\bibitem{Zegers}
R.~G.~T.~Zegers et al.,
Phys.\ Rev.\ Lett. {\bf 91} (2003) 092001.

\bibitem{kff1}
B.-W.~Xiao and X.~Qian,
Eur.\ Phys.\ J.\ A {\bf 15} (2002) 523
[arXiv:hep-ph/0209138].

\bibitem{kff2}
F.~P.~Pereira et al.,
Phys.\ Part.\ Nucl. {\bf 36} (2005) 217
[arXiv:hep-ph/0506032].

\bibitem{kff3}
X.-G.~Wu, and T.~Huang,
JHEP 0804:043, 2008
[arXiv:hep-ph/0803.4229].

\bibitem{Bijnens}
J.~Bijnens and A.~Khodjamirian,
Eur.\ Phys.\ J.\ C {\bf 26} (2002) 67
[arXiv:hep-ph/0206252].

\bibitem{St}
J.~Botts and G.~Sterman,
Nucl.\ Phys.\ B {\bf 325} (1989) 62.

\bibitem{LiSt}
H.-N.~Li and G.~Sterman, 
Nucl.\ Phys.\ B {\bf 381} (1992) 129.

\bibitem{wfs0}
V.~L.~Chernyak and A.~R.~Zhitnitsky, 
JETP Lett. {\bf 25} (1977) 510;
Yad.\ Fiz. {\bf 31} (1980) 1053;
Nucl.\ Phys.\ B {\bf 201} (1982) 492;
Sov.\ J.\ Nucl.\ Phys. {\bf 38} (1983) 775;
Nucl.\ Phys.\ B {\bf 246} (1984) 52.
Phys.\ Rep. {\bf 112} (1984) 173.

\bibitem{wfs1}
V.~L.~Chernyak, A.~R.~Zhitnitsky and I.~R.~Zhitnitsky
Nucl.\ Phys.\ B {\bf 204} (1982) 477.

\bibitem{wfs2}
V.~L.~Chernyak, A.~R.~Zhitnitsky and V.~G.~Serbo,
JETP Lett. {\bf 26} (1977) 594;
Sov.\ J.\ Nucl.\ Phys. {\bf 31} (1980) 552.

\bibitem{wfs3}
B.~V.~Geshkenbeim and M.~V.~Terentyev,
Phys.\ Lett.\ B {\bf 117} (1982) 243;
Sov.\ J.\ Nucl.\ Phys. {\bf 39} (1984) 554;
Sov.\ J.\ Nucl.\ Phys. {\bf 39} (1984) 873.

\bibitem{wfs4}
C.~S.~Huang,
Commun.\ Theor.\ Phys. {\bf 2} (1983) 1265.

\bibitem{wfs5}
M.~Gari and N.~G.~Stefanis,
Phys.\ Lett.\ B {\bf 175} (1986) 462.

\bibitem{wfs6}
Z.~Dziembowski and L.~Mankiewicz,
Phys.\ Rev.\ Lett. {\bf 58} (1987) 2175.

\bibitem{wfs7}
T.~Huang and Q.~X.~Sheng,
Z.\ Phys.\ C {\bf 50} (1991) 139.

\bibitem{wfs8}
F.-G.~Cao, T.~Huang and B.~Q.~Ma,
Phys.\ Rev.\ D {\bf 53} (1996) 6582.

\bibitem{wfs9}
S.~Brodsky and G.~F.~de Teramond,
[arXiv:hep-ph/0804.3562].

\bibitem{DAs1}
V.~M.~Braun and I.~E.~Filyanov,
Z.\ Phys.\ C {\bf 44} (1989) 157.

\bibitem{DAs2}
P.~Ball,
JHEP {\bf 9901} (1999) 010.

\bibitem{DAs3}
P.~Ball and M.~Boglione,
Phys.\ Rev.\ D {\bf 68} (2003) 094006.

\bibitem{DAs4}
P.~Ball, V.~M.~Braun and A.~Lenz,
JHEP {\bf 0605} (2006) 004
[arXiv:hep-ph/0603063].

\bibitem{DAs5}
A.~Khodjamirian, Th.~Mannel and M.~Melcher,
Phys.\ Rev.\ D {\bf 68} (2003) 114007;
Phys.\ Rev.\ D {\bf 70} (2004) 094002
[arXiv:hep-ph/0407226].

\bibitem{frozen1}
C.-R.~Ji and F.~Amiri,
Phys.\ Rev.\ D {\bf 42} (1990) 3764.

\bibitem{frozen21}
J.~M.~Cornwall,
Phys.\ Rev.\ D {\bf 26} (1982) 1453.

\bibitem{frozen22}
C.-R.~Ji, A.~F.~Sill and R.~M.~Lombdar-Nelson,
Phys.\ Rev.\ D {\bf 36} (1987) 165.

\bibitem{VMD1}
W.~G.~Holladay,
Phys.\ Rev. {\bf 101} (1956) 1198.

\bibitem{VMD2}
W.~R.~Frazer and J.~R.~Fulco,
Phys.\ Rev.\ Lett. {\bf 2} (1959) 365;
Phys.\ Rev. {\bf 117} (1960) 1609.

\bibitem{VMD3}
J.~Sakurai,
{\it Currents and Mesons} (University of Chicago, Chicago, 1969).

\bibitem{Anant}
B.~Ananthanarayan and S.~Ramanan,
Eur.\ Phys.\ J.\ C {\bf 54} (2008) 461
[arXiv:hep-ph/0801.2023].

\bibitem{asy}
S.~J.~Brodsky and G~.R~.~Farrar,
Phys.\ Rev.\ Lett. {\bf 31} (1973) 1153;
Phys.\ Rev.\ D {\bf 11} (1975) 1309.

\bibitem{FJ}
F.~R.~Farrar and D.~R.~Jackson,
Phys.\ Rev.\ Lett. {\bf 43} (1979) 246.

\bibitem{pff01}
V.~A.~Nesterenko and A.~V.~Radyushkin, 
Phys.\ Lett.\ B {\bf 115} (1982) 410.

\bibitem{pff02}
V.~V.~Anisovich, D.~I.~Melikhov and V.~A.~Nikonov,
Phys.\ Rev.\ D {\bf 52} (1995) 5295;
Phys.\ Rev.\ D {\bf 55} (1997) 2918.

\bibitem{pff03}
A.~P.~Bakulev, A.~V.~Radyushkin and N.~G.~Stefanis,
Phys.\ Rev.\ D {\bf 62} (2000) 113001.

\bibitem{pff04}
V.~Braguta, W.~Lucha and D.~Melikhov,
Phys.\ Lett.\ B {\bf 661} (2008) 354
[arXiv:hep-ph/0710.5461].

\bibitem{pff11}
C.~R.~Ji and S.~Cotanch,
Phys.\ Rev.\ D {\bf 21} (1990) 2319.

\bibitem{pff12}
T.~Gousset and B.~Pire,
Phys.\ Rev.\ D {\bf 51} (1995) 15;
[arXiv:hep-ph/9403293];
Proceedings of the {\it ELFE Summer School on Confinement physics, Cambridge 1995}, p.111-143
[arXiv:hep-ph/9511274].

\bibitem{pff13}
C.-R.~Ji, A.~Pang and A.~Szczepaniak,
Phys.\ Rev.\ D {\bf 52} (1995) 4038.

\bibitem{pff14}
J.~P.~B.~C.~de~Melo et al.,
Phys.\ Rev.\ C {\bf 59} (1999) 2278;
Nucl.\ Phys.\ A {\bf 707} (2002) 399

\bibitem{pff15}
H.-M.~Choi and C.-R.~Ji,
Phys.\ Rev.\ D {\bf 74} (2006) 093010
[arXiv:hep-ph/0608148];
[arXiv:hep-ph/0803.2604].

\bibitem{pff2}
F.-G.~Cao, Y.-B.~Dai and C.-S.~Huang,
Eur.\ Phys.\ J.\ C {\bf 11} (1999) 501
[arXiv:hep-ph/9711203].

\bibitem{pff3}
C.~Coriano, H.-N.~Li and C.~Savkli, 
JHEP 9807:008, 1998
[arXiv:hep-ph/9805406].

\bibitem{pff4}
V.~M.~Braun, A.~Khodjamirian and M.~Maul,
Phys.\ Rev.\ D {\bf 61} (2000) 073004
[arXiv:hep-ph/9907495].

\bibitem{pff5}
N.~G.~Stefanis, W.~Schroers and H.-Ch.~Kim,
Eur.\ Phys.\ J.\ C {\bf 18} (2000) 137.

\bibitem{pff6}
A.~P.~Bakulev, et al.,
Phys.\ Rev.\ D {\bf 70} (2004) 033014
[arXiv:hep-ph/0405062].

\bibitem{pff7}
Z.-T.~Wei and M.-Z.~Yang,
Phys.\ Rev.\ D {\bf 67} (2003) 094013.

\bibitem{pff8}
T.~Huang, X.-G.~Wu and  X.-H.~Wu,
Phys.\ Rev.\ D {\bf 70} (2004) 053007
[arXiv:hep-ph/0404163].

\bibitem{pff9}
T.~Huang and X.-G.~Wu,
Phys.\ Rev.\ D {\bf 70} (2004) 093013
[arXiv:hep-ph/0408252].

\bibitem{Lattice1}
F.~D.~R.~Bonnet et al. [Lattice Hadron Physics Collaboration],
Phys.\ Rev.\ D {\bf 72} (2005) 054506
[arXiv:hep-lat/0411028].

\bibitem{Lattice2}
S.~Hashimoto et al. [JLQCD Collaboration],
PoS {\bf LAT2005}, (2006) 336
[arXiv:hep-lat/0510085].

\bibitem{Lattice3}
D.~Br{\"o}mmel et al., 
PoS {\bf LAT2005}, (2006) 360
[arXiv:hep-lat/0509133];
Eur.\ Phys.\ J.\ C {\bf 51} (2007) 335
[arXiv:hep-lat/0608021].

\bibitem{Lattice4}
P.-H.~J.~Hsu and G.~T.~Fleming
[arXiv:hep-lat/0710.4538].

\bibitem{crit4}
A.~P.~Bakulev and A.~V.~Radyushkin,
Phys.\ Lett.\ B {\bf 271} (1991) 223.

\bibitem{Brodsky}
G.~P.~Lepage and S.~J.~Brodsky,
Phys.\ Lett.\ B {\bf 87} (1979) 359;
Phys.\ Rev.\ Lett. {\bf 43} (1979) 545;
Phys.\ Rev.\ D {\bf 22} (1980) 2157;
Perturbative Quantum Chromodynamics, A.~H.~Mueller ed., p.93, World 
Scientific, Singapore 1989;
S.~J.~Brodsky, in Proceedings of the {\it Quantum Chromodynamics} Workshop, La 
Jolla, California, 1978.

\bibitem{Efremov}
A.~V.~Efremov, and A.~V.~Radyushkin, 
Phys.\ Lett.\ B {\bf 94} (1980) 245;
Theor.\ Math.\ Phys. {\bf 42} (1980) 97.

\bibitem{Radyushkin}
A.~V.~Radyushkin,
[arXiv:hep-ph/0410276].

\bibitem{Mueller}
A.~Duncan and A.~H.~Mueller,
Phys.\ Rev.\ D {\bf 21} (1980) 1636.

\bibitem{condensate1}
S.~V.~Mikhailov and A.~V.~Radyushkin,
JETP Lett. {\bf 43} (1986) 712;
Sov.\ J.\ Nucl.\ Phys. {\bf 49} (1989) 494;
Phys.\ Rev.\ D {\bf 45} (1992) 1754.

\bibitem{condensate2}
A.~P.~Bakulev and A.~V.~Radyushkin,
Phys.\ Lett.\ B {\bf 271} (1991) 223.

\bibitem{condensate3}
A.~P.~Bakulev and S.~V.~Mikhailov,
Z.\ Phys.\ C {\bf 68} (1995) 451.

\bibitem{crit0}
B.~L.~Ioffe and A.~V.~Smilga,
Phys.\ Lett.\ B {\bf 114} (1982) 353.

\bibitem{crit1}
N.~Isgur and C.~H.~Llewellyn Smith,
Phys.\ Rev.\ Lett. {\bf 52} (1984) 1080;
Nucl.\ Phys.\ B {\bf 317} (1989) 526;
Phys.\ Lett.\ B {\bf 217} (1989) 535.

\bibitem{crit2}
A.~V.~Radyushkin,
Acta.\ Phys.\ Polonica\ B {\bf 15} (1984) 403;
Nucl.\ Phys.\ A {\bf 527} (1991) 153C;
Nucl.\ Phys.\ A {\bf 532} (1991) 141.

\bibitem{crit3}
O.~C.~Jacob and L.~S.~Kisslinger,
Phys.\ Rev.\ Lett. {\bf 56} (1986) 225.

\bibitem{crit5}
R.~Jacob and P.~Kroll,
Phys.\ Lett.\ B {\bf 315} (1993) 463.

\bibitem{crit6}
J.~Bolz et al.,
Z.\ Phys.\ C {\bf 66} (1995) 267.

\bibitem{crit7}
V.~M.~Braun and I.~Halperin, 
Phys.\ Lett.\ B {\bf 328} (1994) 457.

\bibitem{crit8}
B.~Chibisov and A.~R.~Zhitnitsky,
Phys.\ Rev.\ D {\bf 52} (1995) 5373.

\bibitem{crit9}
D.~Melikhov,
Phys.\ Rev.\ D {\bf 53} (1996) 2460;
Eur.\ Phys.\ J. direct C {\bf 4} (2002) 2
[arXiv:hep-ph/0110087].

\bibitem{Descotes}
S.~Descotes and C.~T.~Sachrajda,
Nucl.\ Phys.\ B {\bf 625} (2002) 239.

\bibitem{BHL}
S.~J.~Brodsky, G.~P.~Lepage and T.~Huang,
Invited Talk at the Banff Summer Institute on Particle and Fields (1981),
A.~Z.~Capri and A.~N.~Kamal, ed., p.143, Plenum Press, New York 1983.

\bibitem{SVZ}
M.~A.~Shifman, A.~I.~Vainshtein and V.~I.~Zakharov,
Nucl.\ Phys.\ B {\bf 147} (1979) 385.

\bibitem{Shirkov}
D.~V.~Shirkov and I.~L.~Solovtsov, 
Phys.\ Rev.\ Lett. {\bf 79} (1997), 1209;
Phys.\ Lett.\ B {\bf 442} (1998), 344;
Theor.\ Math.\ Phys. {\bf 120} (1999) 1220.

\bibitem{analytic}
N.~G.~Stefanis, W.~Schroers and H.-Ch.~Kim,
Phys.\ Lett.\ B {\bf 4429} (1999) 299.
[arXiv:hep-ph/9807298].

\bibitem{LiBD}
H.-N.~Li,
Phys.\ Rev.\ D {\bf 52} (1995) 3958.

\bibitem{Beneke}
M.~Beneke and Th.~Feldmann,
Nucl.\ Phys.\ B {\bf 592} (2001) 3.

\bibitem{Wei}
Z.~T.~Wei and M.~Z.~Yang,
Nucl.\ Phys.\ B {\bf 642} (2002) 263.

\bibitem{Sanda}
T.~Kurimoto, H.-N.~Li and A.~I.~Sanda,
Phys.\ Rev.\ D {\bf 65} (2002) 014007.

\bibitem{Li}
H.-N.~Li,
Phys.\ Rev.\ D {\bf 66} (2002) 094010.


\end{thebibliography}
\end{document}